\documentclass[aps,prl,amsmath,twocolumn,superscriptaddress,floatfix]{revtex4-1}
\usepackage{amsmath}
\usepackage{amssymb}
\usepackage{graphicx}
\usepackage{bbold}
\usepackage{float}
\usepackage[colorlinks=true,urlcolor=blue,anchorcolor=blue,linkcolor=blue,citecolor=blue,breaklinks=true]{hyperref}
\usepackage{adjustbox}
\makeatletter
\usepackage{booktabs,multirow}
\usepackage[table]{xcolor}
\newcommand{\beginsupplement}{%
	\setcounter{table}{0}
	\renewcommand{\thetable}{T\arabic{table}}%
	\setcounter{figure}{0}
	\renewcommand{\thefigure}{S\arabic{figure}}%
	\setcounter{section}{0}
	\renewcommand{\thesection}{\Alph{section}}%
	\setcounter{section}{0}
	\renewcommand{\thesubsection}{\alph{subsection}}%
	\setcounter{subsection}{0}
	\renewcommand{\thesubsection}{S\arabic{section}.\arabic{subsection}}%
	\setcounter{equation}{0}
	\renewcommand{\theequation}{\Roman{equation}}%
	\setcounter{enumiv}{0}
	\renewcommand{\theenumiv}{S\arabic{enumiv}}%
}
\makeatother

\def\To{T_c^{\rm{\tiny onset}}}

\begin{document}
	
	\title{Enhanced two-component superconductivity in CoSi$_2$/TiSi$_2$ heterojunctions}
	
	\author{Shao-Pin Chiu}
	\thanks{These authors contributed equally to this work.}
	\affiliation{Department of Electrophysics, National Yang Ming Chiao Tung University, Hsinchu 30010, Taiwan}
	\affiliation{Center for Emergent Functional Matter Science, National Yang Ming Chiao Tung University, Hsinchu 30010, Taiwan}
	
	\author{Vivek Mishra}
	\thanks{These authors contributed equally to this work.}
	\affiliation{Kavli Institute for Theoretical Sciences, University of Chinese Academy of Sciences, Beijing 100190, China}
	
	\author{Yu Li}
	\affiliation{Kavli Institute for Theoretical Sciences, University of Chinese Academy of Sciences, Beijing 100190, China}
	
	\author{Fu-Chun Zhang}
	\affiliation{Kavli Institute for Theoretical Sciences, University of Chinese Academy of Sciences, Beijing 100190, China}
	\affiliation{CAS Center for Excellence in Topological Quantum Computation, University of Chinese Academy of Sciences, Beijing 100190, China}
	\affiliation{{HKU-UCAS Joint Institute of Theoretical and Computational Physics at Beijing, University of Chinese Academy of Sciences, Beijing 100190, China}}
	
	\author{Stefan Kirchner}
	\email{Corresponding author: stefan.kirchner@correlated-matter.com}
	\affiliation{Department of Electrophysics, National Yang Ming Chiao Tung University, Hsinchu 30010, Taiwan}
	\affiliation{Center for Emergent Functional Matter Science, National Yang Ming Chiao Tung University, Hsinchu 30010, Taiwan}
	
	\author{Juhn-Jong Lin}
	\email{Corresponding author: jjlin@nycu.edu.tw}
	\affiliation{Department of Electrophysics, National Yang Ming Chiao Tung University, Hsinchu 30010, Taiwan}
	\affiliation{Center for Emergent Functional Matter Science, National Yang Ming Chiao Tung University, Hsinchu 30010, Taiwan}

	\date{\today}
	\begin{abstract}
		We report  enhanced two-component superconductivity in (CoSi$_2$/Si)/TiSi$_2$ superconductor/normal-metal (S/N) heterojunctions. An enhanced superconducting transition temperature $\To$  {about twice that} of CoSi$_2$  and an upper critical field  $\sim$\,20 times bigger than that of epitaxial CoSi$_2$/Si films were found. The {tunneling} spectra of three-terminal S/N junctions show pronounced zero-bias conductance peaks (ZBCPs) that signify penetration  of odd-frequency, spin-triplet and even-parity Cooper pairs in TiSi$_2$ from triplet dominant {pairing in} {CoSi$_2$/Si} {driven by symmetry reduction at the {CoSi$_2$/Si} interface}. Both the enhancement of $\To$ value and the ZBCPs are found to be more pronounced if TiSi$_2$ is made more diffusive.
	\end{abstract}
	\maketitle
	
\begin{figure}[t!]
		\centering
		\includegraphics[width= 0.5\textwidth]{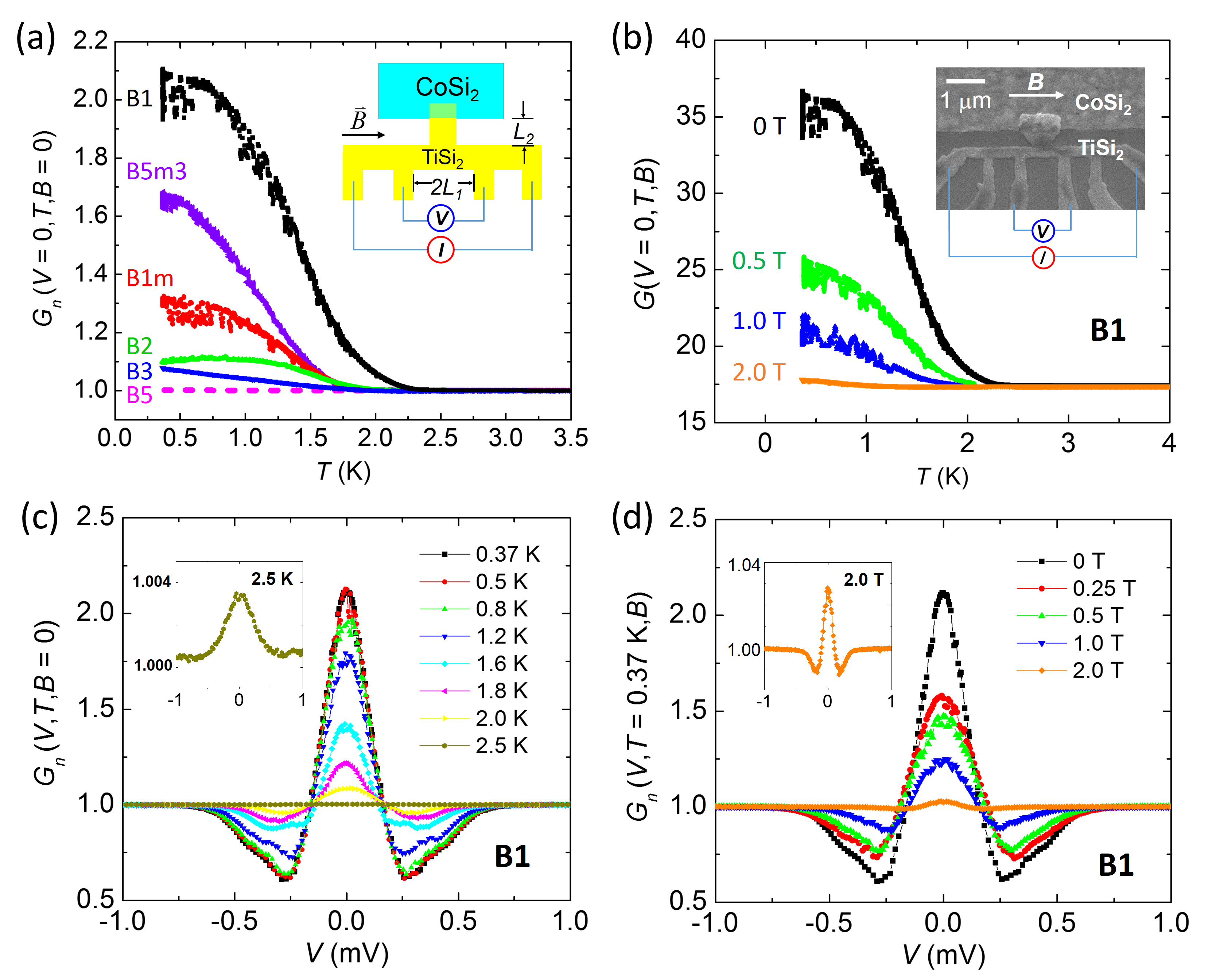} 
		\caption{
			(a) Temperature dependence of  $G_n(0,T,0)$ in six  devices, as indicated. Inset: a schematic CoSi$_2$/TiSi$_2$ T-shaped proximity structure with a 4-probe configuration. 2$L_1$ is the distance separating the voltage electrodes. $L_2$ is the length of the TiSi$_2$ segment connecting to CoSi$_2$/Si.  (b)--(d) $G_n(V,T,0)$ of the as-grown device B1. (b) Zero-bias $G(0,T,B)$ in several $B$ fields. Inset: an SEM image of the device. (c)  $G_n(V,T,0)$ at several $T$ values. Inset: a zoom-in of the 2.5-K curve. (d) $G_n(V,0.37\,{\rm K},B)$ in several $B$ fields. Inset: a zoom-in of the 2.0-T curve. 
		}
		\label{fig1}
	\end{figure}

	The physics of novel quantum states of matter and in particular non-conventional superconducting states have been central of recent condensed matter research. This development is driven by scientific interest and technological prospects. Notably, the ability to engineer and manipulate entangled quantum states in dedicated devices holds promise  for realizing topological quantum computation \cite{Read.00,Nayak.08,Qi.11,Alicea.12}.
	A major challenge is the identification of systems that incorporate both aspects of generation and manipulation of such quantum states at the microscopic scale. This involves the identification of non-trivial pairing states.
	A three terminal {T}-shaped proximity structure [Fig. \ref{fig1}(a)] had been proposed to aid  distinguishing  triplet from singlet superconductors \citep{Asano2007}.  This {T}-shaped proximity structure detects triplet pairing through a zero-bias conductance peak (ZBCP) that results from the generation of odd-frequency spin-triplet even-parity pairs in the diffusive normal metal (DN) part of the superconductor/normal metal (S/N) interface \cite{Tanaka2005,*Tanaka2005_ERR,TanakaGolubov2007}. Yet, the fabrication of such T-shaped structures has proven difficult for many materials \cite{Courtois1999,Mackenzie2017}. Moreover, interfaces and electronic confinement that appear to be inherent to such devices can also give rise to such intricate phenomena like interface superconductivity \cite{Bozovic.11}.
	
	The successful fabrication of high-quality CoSi$_2$/TiSi$_2$ T-shaped proximity structures on silicon was recently reported \cite{Chiu.21,Chiu.21jjap}. CoSi$_2$ is a  superconductor that is widely used in the semiconductor industry with a  transition temperature ($T_c^{\text{\tiny bulk}}$) of $1.3$K 
	\cite{CoSi2SC,CoSi2SC_b,CoSi2SH,CoSi2ES}. Interestingly, Chiu {\itshape et al.} established the existence of triplet pairing in these junctions {\itshape via} the {anomalous proximity effect} (APE) that leads to  a ZBCP \cite{Chiu.21}.
	A possible way of understanding these findings in terms of the symmetry reduction brought about by the underlying dielectric  substrate was proposed in Ref.\ \cite{Mishra.21}. This  is in line with the finding that the accompanying interface between CoSi$_2$ and the Si(100) substrate \cite{Chiu.17} gives rise to a spin-orbit coupling (SOC) which  exceeds the bulk CoSi$_2$ superconducting energy gap $\Delta_0$ by a factor $\sim$\,30 \cite{Chiu.21}. Therefore, parity no longer remains a good quantum number and that reflects in the gap structure $\hat{\Delta}= \left( \Delta_s \mathbb{1}  + \Delta_t \mathbf{d} \cdot \boldsymbol{\sigma} \right) i\sigma_y$, which is a combination of   singlet $\Delta_s$ and  triplet $\Delta_t$ components. $\hat{\Delta}$ is a matrix in spin-space, and $\mathbf{d}$ is the $d$-vector of the triplet pairing, $\boldsymbol{\sigma}, \sigma_y$ denote the Pauli matrices, and $\mathbb{1}$ is the identity matrix in spin-space. Highest $T_c$ is realized when the $\mathbf{d}$ is along the SOC field \cite{GorkovRashba,Frigeri2004}, which is $ \hat{z}\times \mathbf{k}$ in the present case due to broken inversion symmetry along the $\hat{z}$ axis. Here $\mathbf{k}$ is the momentum vector.
	
	In this Letter, we report  superconductivity 
	in CoSi$_2$/TiSi$_2$ heterostructures with an onset temperature ($\To$) of roughly twice of $T_c^{\text{\tiny bulk}}$. The presence of superconductivity in the junction is inferred through the APE. Our in-depth comparison with  Ref.\ \cite{Mishra.21} indicates a mixture of singlet and triplet pairing channels with a dominant triplet component which drives the {APE}. A Ginzburg-Landau (GL) analysis is presented  that captures the $\To$ enhancement and provides a modeling of the system.  Our analysis also sheds light on the  roles played by the different interfaces forming the  CoSi$_2$/TiSi$_2$ heterostructures. 
	
	TiSi$_2$ is a DN and remains metallic down to $T=50$ mK \cite{Chiu.21}. With respect to the CoSi$_2$/TiSi$_2$ proximity structures, it will be important that TiSi$_2$ can exist in a base-centered phase (C49) and a face-centered phase (C54) \cite{Mattheiss1989,Ekman1998}.
	The C49 phase is known to contain large amounts of stacking faults \cite{MaAllen1994}.  It has an order of magnitude larger resistivity [$\rho$(300\,K) $\simeq$ (100--200) $\mu \Omega$ cm] than the C54 phase [$\rho$(300\,K) $\simeq$ (17--25) $\mu \Omega$ cm] \cite{SupMat}. This {inherent material property} allows us to investigate the characteristics of the CoSi$_2$/TiSi$_2$ proximity structure in terms of the diffusive properties of the normal-metal component. 
	
	Surprisingly, in a number of devices, $\To$ is significantly enhanced ($\To > 2$\,K). The enhanced $\To$ value is not only higher than those in the T-shaped structures studied in Ref. \cite{Chiu.21} ($\To \leq 1.4$\,K), but also higher than the  {$T_c$} value ($\approx$ 1.5\,K) of epitaxial CoSi$_2$/Si films \cite{Chiu.17}. Moreover, the normalized amplitude of ZBCP is strongly enhanced up to $\approx$\,210\% with respect to the (normalized) conductance of the normal state. In Fig. \ref{fig1}(a), the inset depicts a schematic T-shaped structure with a 4-probe configuration for measuring the proximity effect in TiSi$_2$, {which is $\approx$\,125-nm thick and typically (0.3$\pm$0.1)-$\mu$m wide}. {(The thickness of the  CoSi$_2$/Si films is $\approx$\,105 nm in this work.)} The main panel shows the normalized zero-bias conductance $G_n(V=0,T,B=0)$ as a function of $T$ for { six} devices, where $G_n$ denotes the differential conductance, $G(V,T,B)\equiv dI(V,T,B)/dV$, normalized to its normal-state value, where $I$ is the current, $V$ is the bias voltage, $T$ is the temperature, and $B$ is the magnetic field.
	
	In Fig.\ \ref{fig1}(a), devices B1, B2, B3 and B5 were measured as grown. Device B1 (B5) then underwent one (three) thermal cycling to room temperature and cooled down again for a second- (fourth-)run measurement [the device is subsequently relabeled B1m (B5m3)]  \cite{SupMat}. The $G_n(0,T,0)$ of device B1 (B1m) increases with decreasing $T$ below an onset temperature $\To = 2.94$\,K (2.84\,K), reaching 212\% (131\%) at 0.365\,K. There are visible conductance fluctuations which are probably induced by dynamical structural defects in this particular device \cite{Yeh.17}. The $G_n(0,T,0)$ in device B2 shows a non-monotonic $T$ dependence, increasing below 2.33\,K and reaching a maximum at $\sim$\,0.75\,K, then followed by a small decrease to $\sim$\,110\% at 0.37\,K. The $G_n(0,T,0)$ in device B3 shows a monotonic increase below 2.14\,K, reaching $\sim$\,106\% at 0.37\,K.  $G_n(0,T,0)$ of device B5 is relatively small compared to that of device B5m3. 
	Figure \ref{fig1}(a) demonstrates that the magnitude of the {APE} is strongly influenced by the high resistivity, {\itshape i.e.}, low electron diffusivity, of the C49 phase.
	
	Figure \ref{fig1}(b) depicts the $G(0,T,B)$ of device B1 in several $B$ fields. In  this work, the $B$ field was applied in the CoSi$_2$/Si plane and parallel to the S/N interface.
	The $G(0,T,B)$ is gradually suppressed with increasing $B$ but a small proximity effect is still visible in $B=2$\,T which is much higher than the in-plane upper critical field ($\leq 0.12$\,T) of CoSi$_2$/Si films \cite{Chiu.21}. Figure \ref{fig1}(c) depicts $G_n(V,T,0)$ of the same device at several $T$ values and in $B=0$. ZBCPs are notable, which are gradually suppressed with increasing $T$. The ZBCP persists up to at least 2.5\,K ({see a zoom-in in the} inset). Figure \ref{fig1}(d) shows $G_n(V,0.37\,{\rm K},B)$ of the same device in several $B$ fields and at $T=0.37$\,K. While the ZBCP is gradually suppressed with increasing $B$, it persists up to at least 2\,T (Fig. \ref{fig1}d inset).
	
	\begin{figure}
		\begin{center}
			\includegraphics[width=0.99\linewidth]{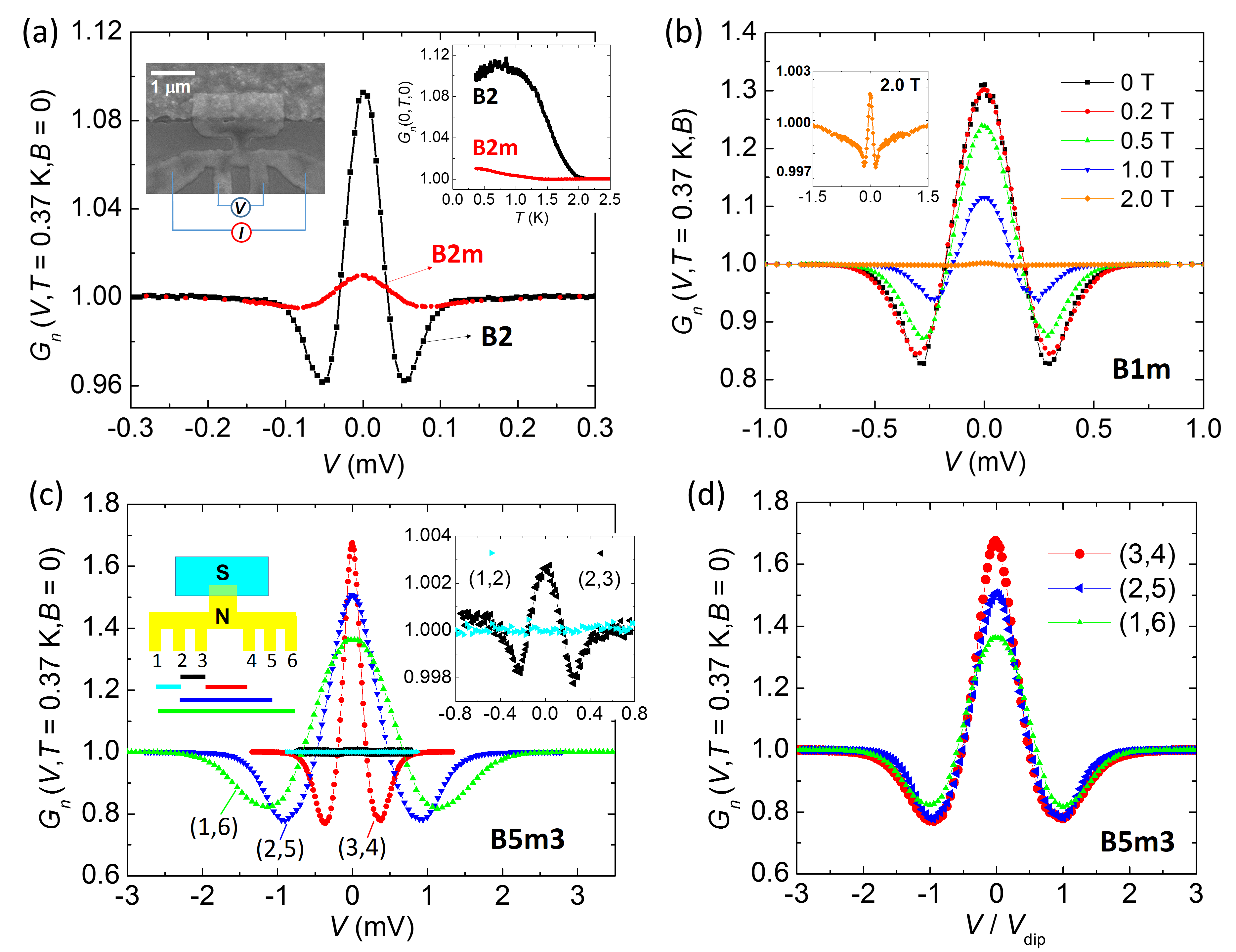}
		\end{center}\vspace{-1em}
		\caption{
			(a)  $G_n(V,0.37\,{\rm K},0)$ of devices B2 and B2m. Left inset: an SEM image of device B2. Right inset: $G_n(0,T,0)$ for the two devices. (b) $G_n(V,0.37\,{\rm K},B)$ of device B1m. Inset: a zoom-in of the 2.0-T curve.
			(c) $G_n(V,0.365\,{\rm K},0)$ of device B5m3 for various voltage-electrode (VE) pairs indicated by the color bars in left inset. $I$ was always applied using the electrode pair (1,6). Right inset: a zoom-in of $G_n(V,0.365\,{\rm K},0)$ for VE pairs (1,2) and (2,3). (d) $G_n(V,0.365\,{\rm K},0)$ for VE pairs (3,4), (2,5), and (1,6) as a function of independent variable $V/V_{\rm dip}$, where $\pm V_{\rm dip}$ is the bias voltages at which the side dips occur in each VE pair configuration.
		}%
		\label{fig2}
	\end{figure}

	To test the robustness of the ZBCPs and the effect of thermal cycling on the {APE}, Fig. \ref{fig2}(a) shows the $G_n(V,0.37\,{\rm K},0)$ of B2 and B2m at $T=0.37$\,K and in $B$ = 0. (B2m denotes the second-run measurement of B2). The amplitude of ZBCP of B2m is notably reduced from that of B2. Nevertheless, it remains readily detectable. The right inset shows that the $\To$ in $G_n(0,T,0)$ is reduced from 2.33\,K in B2 to 1.42 K in B2m.
	
	We use device B1m to  demonstrate that the {APE} persists up to  large $B$ fields. Figure \ref{fig2}(b) shows $G_n(V,0.37\,{\rm K},B)$ of device B1m in several $B$ fields and at $T=0.37$\,K. Although the amplitudes are smaller compared with those in Fig. \ref{fig1}(d), the ZBCPs are robust against thermal cycling. 
	The inset shows a  zoom-in for the 2.0-T curve, where  the ZBCP is still not completely suppressed.
	
	Thus, the superconductivity enhancement is robust and closely correlated with the diffusivity of TiSi$_2$. We have thermally cycled device B5 three times to further investigate the {APE}. In addition, we use this device to illustrate that the ZBCPs only occur in the vicinity of the CoSi$_2$/TiSi$_2$ interface. The left inset of Fig.\ \ref{fig2}(c) shows a schematic S/N {T}-shaped device with six submicron electrodes attaching N.
	The main panel of Fig. \ref{fig2}(c) shows the $G_n(V,0.365\,{\rm K},0)$ for different voltage-electrode (VE) pairs which define different segments of the TiSi$_2$ component. In all cases, {$I$} was applied through the outermost electrode pair (1,6). Figure \ref{fig2}(c) reveals large amplitudes of ZBCP measured with the VE pairs (3,4), (2,5) and (1,6). In contrast, the right inset shows a zoom-in of the small ZBCPs for the VE pairs (1,2) and (2,3), located away from the S/N interface.  In fact, the $G_n$ curve for the VE pair (1,2) is flat, indicating a complete absence of the {APE}. These results provide unambiguous evidence that the ZBCP must arise from the penetration of Cooper pairs through the CoSi$_2$/TiSi$_2$ interface, as theoretically predicted for triplet superconductivity \cite{Asano2007}. 
	
	Figure \ref{fig2}(d) shows  $G_n(V,0.365\,{\rm K},0)$ for VE pairs (3,4), (2,5) and (1,6) vs.\  $V/V_{\rm dip}$, where $\pm V_{\rm dip}$ are the voltages where the two side dips occur in each VE pair configuration. The lineshapes of all three $G_n$ curves are similar, while the amplitudes of ZBCP decrease with increasing VE pair separation, as expected. When the VE pair separation is large, no Cooper pairs can diffuse to those TiSi$_2$ regimes far away from the S/N interface. Thus, those TiSi$_2$ regimes contribute a finite resistance, leading to a reduced ZBCP height.
	
	\begin{figure}
		\includegraphics[width=1.0\linewidth]{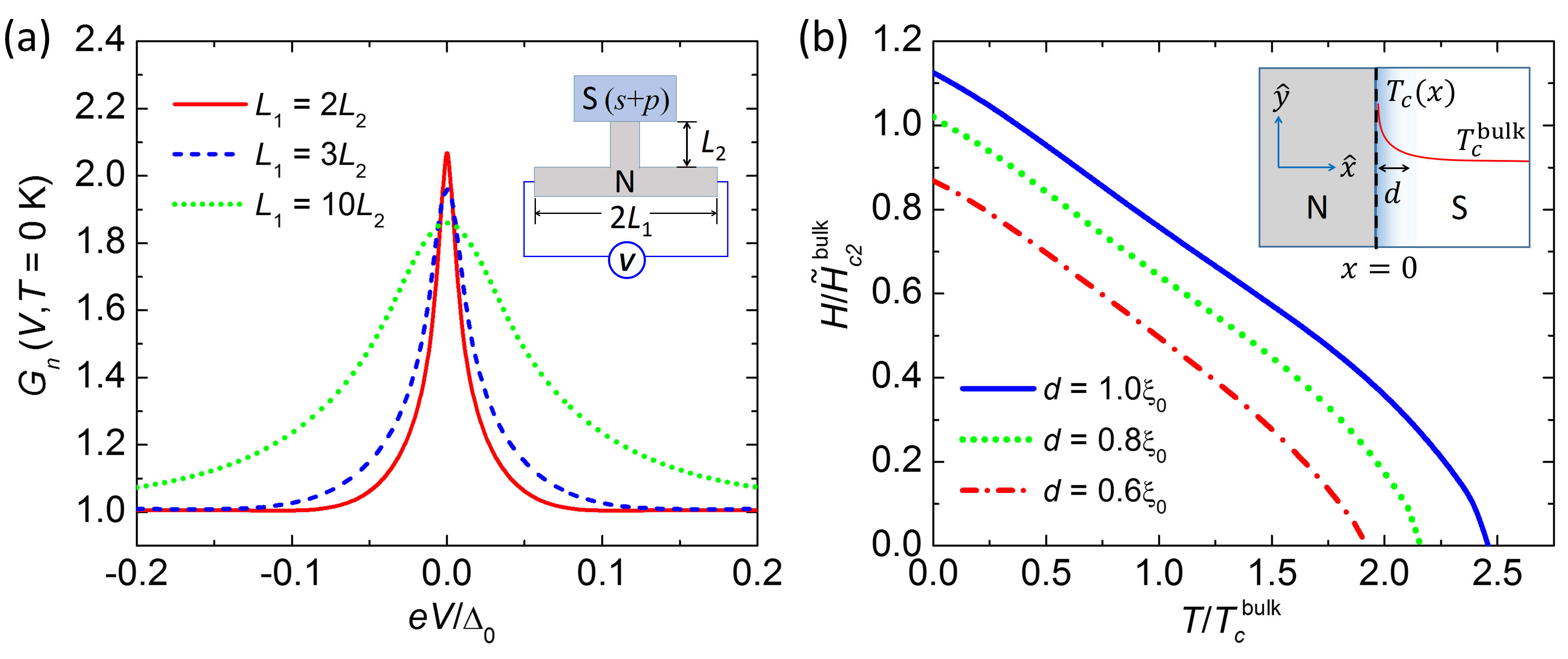}
		\caption{(a) $G_n(V,0,0)$ for {a} T-shaped junction {calculated via the the quasi-classical theory \cite{Asano2007,Mishra.21}} for a $s+p$ superconductor for various values of $L_1$.{ The triplet gap $\Delta_t=2\Delta_0/\sqrt{5}$, the singlet gap $\Delta_s=\Delta_0/\sqrt{5}$, $Z=2$, and electron dephasing rate $\hbar/\tau_{\varphi}=0.05\Delta_0$.} The value of $L_2$ 
			{remains} fixed, $R_N/R_B = 100 L_1/L_2$ and $\Delta_0$ is $20 \hbar D/L_2^2$, where $D$ is the diffusion constant for {DN}.
			{Inset: The junction geometry is shown.} (b) The $H \mbox{-} T$ phase diagram {based on GL theory} for the emergent superconductivity due to enhanced superconducting correlations near the {S/N} interface for several values of $d$ in units of $\xi_0$.  $T$ and $H$ are expressed in units of $T_c^{\rm bulk}$ and  $\tilde{H}^{\rm bulk}_{c2}$, respectively \cite{SupMat}.
			Inset: The S/N interface region is illustrated with the effective interface width `$d$' denoted. A red curve showing enhanced $T_c(x)$ near the interface is a schematic illustration of the model.}
		\label{Fig:T1}
	\end{figure}
	
	We apply a  quasi-classical description of the CoSi$_2$/TiSi$_2$ three-terminal heterojunctions  \cite{Asano2007,Mishra.21} and consider a SOC induced $s+p$ pairing state with a dominant triplet component \cite{Mishra.21}. 
	Within the formulation of the circuit theory \cite{Nazarov1999,Tanaka2003bc,Tanaka2004,*Tanaka2004_ERR}, an insulating  barrier is expressed as a delta function [$Z E_F k_F^{-1} \delta(x)$] at  the {S/N} interface.
	The weight of the barrier is expressed in terms of a dimensionless parameter $Z$, Fermi energy $E_F$, and  the Fermi wavenumber $k_F$. 
	A higher ratio of $R_N/R_B$, {\itshape i.e.},  resistance of the normal-metal component ($R_N$) over  barrier resistance ($R_B$), results in  {an APE} over a broader energy range and thus in an increase in the FWHM of the ZBCP \cite{Asano2007}.  The zero-bias value itself, however, is (roughly) independent of $R_N$/$R_B$ if electron dephasing is ignored in the DN component of the junction. 
	
	In Fig. \ref{Fig:T1}(a), we keep  {$R_B$},  $\Delta_t$, $\Delta_s$ and $Z$ fixed, and vary $L_1$ according to the experiments as shown in Fig. \ref{fig2}(c). {Importantly, when VE configuration is changed, the S/N interface properties remain fixed but as $L_1$ changes $R_N$ changes likewise.
		We find that the ZBCP becomes broader, in qualitative agreement with Fig. \ref{fig2}(c).   
		While this variation of FWHM with $L_1$ agrees with the experiment, the inclusion of a small amount of electron dephasing is required to reproduce the experimental behavior of $G_n(V=0)$, \textit{ i.e.}, a suppression of  $G_n(V=0)$ with increasing $L_1$.
		In diffusive metals, the existence of a finite electron dephasing rate in low-$T$ regimes have been reported for long, but its microscopic origin(s) are yet to be fully identified \cite{Lin_2002}.
		Magnetic impurities as a source of dephasing however appears unlikely as their presence in the heterojunction has been ruled out \cite{Chiu.21}.}
	{For the C49 phase, $R_N$ is higher than in the C54 phase, which results in a larger ZBCP provided other parameters are kept constant.  Across the devices, the behavior of ZBCP can be interpreted along this line, but typically, several device parameters undergo changes from device to device or upon thermal cycling (section C of \cite{SupMat}).
		Nevertheless, the robust ZBCPs in these devices support the interpretation in terms of the SOC generated two-component superconductivity with a dominant triplet component, which is essential for the ZBCP.}
	
	The enhanced  $\To$ is usually found in T-shaped structures with high-$\rho$ TiSi$_2$. This leads to the important observation that when $\rho$(TiSi$_2$) is reduced by thermal cycling, the enhancement of superconductivity is also reduced, {\itshape i.e.}, the enhanced superconductivity is correlated with some kind of  defect structures. The case of B5 (and B5m3) further supports this conclusion because after thermal cycling B5m3 became more resistive and enhanced superconductivity emerged.
	
	The quasi-classical theory provides a good explanation for the robustness of the ZBCPs in T-shaped junctions,
	but does not
	take into account the enhanced superconductivity found in these devices.
	As the interface quality plays a vital role in the enhancement of superconductivity, we will assume that  somewhere near or at the S/N interface there exists another superconducting phase whose $T_c$ is higher than $T_c^{\rm bulk}$. Its microscopic origin
	could be changes in the electronic structure or changes in the phonon modes, as in  YIr$_2$-Ir and EuIr$_2$-Ir eutectic systems \cite{Mathhias1980,Suhl1980}. Here we focus on a phenomenological description of this effect rather than a microscopic modeling.
	
	We adopt  GL  theory for the singlet-triplet mixed superconducting state to understand this phenomena.  
	The free energy for such a system  reads \cite{Sigrist_Review},
	\begin{eqnarray}
		\mathbf{F} &=& \int d\vec{r} \left[ \mathcal{F}_{0} +  \mathcal{F}_{c} + \mathcal{F}_{n}+\mathcal{F}_{B} + \mathcal{F}_{me}+ \mathcal{F}_{pol} \right]
		\label{Eq:FE}
	\end{eqnarray}
	here $\mathcal{F}_{0} = \sum_{\nu} \big(a_{\nu}(x)|\Phi_{\nu}|^2 + b_\nu |\Phi_{\nu}|^4 + K_\nu |\vec{D}\Phi_{\nu}|^2\big)$ is the usual GL free energy for each individual component, the subscript $\nu=\pm$ represents the two components of the order parameter, and  $\vec{D}$ is $\vec{\nabla}-2\pi i /\Phi_0 \vec{A}$, where $\vec{A}$ is the vector potential, and $\Phi_0$ is the magnetic flux quantum. The coefficient $a_\nu$ is $\alpha_\nu (T-T_{c\nu})$. We assume a clean superconductor and  $K_\pm=K_s$. $\mathcal{F}_{c}=c(x) \left( \Phi_+^* \Phi_- + \Phi_+ \Phi_-^* \right)$ is the coupling between the two components, a negative value of the coefficient $c$ ensures a single transition temperature and  $\mathcal{F}_{n}$  is the free energy density of the non-superconducting state without magnetic field ($H$).
	$\mathcal{F}_B=\vec{\tilde{B}}^2/(8\pi)$ is the contribution from the magnetic field \footnote{$\vec{B}$ denotes the total magnetic field. $\vec{H}$ refers to the external field as common in the GL literature. $H$  has to be identified with $B$ of Figs.\ \ref{fig1} and \ref{fig2}.}.  $\mathcal{F}_{me}=\sum_{\nu} i K_{me,\nu} (\vec{H}\times \hat{z})\cdot \left[ \Phi_\nu^* ( \vec{D}\Phi_\nu) - h.c. \right]$ is a Lifshitz invariant term that leads to a magneto-electric coupling and $\mathcal{F}_{pol}=\sum_{\nu,i} Q_{\nu i}   |\Phi_\nu|^2  H^2_{i}$  determines the effect of superconducting order on spin-polarization, where  $i=x,y,z$ labels  spatial components.  
	
	We take the S/N interface at $x=0$ and assume a homogeneous system along the other directions. We restrict ourselves to the experimental field configuration, \textit{i.e.}, along the interface in the plane ($\hat{y}$).  The coefficients of the quadratic terms are taken to be spatially varying to model the enhanced superconductivity near the interface. This dependence of $a_\pm(x)$ is $\alpha_\pm [T - T_{c\pm} (x)]$, where  $T_{c \pm}(x) = T_{c \pm} \left[ 1 + \eta_\pm \mathrm{sech}\left(\frac{x}{d}\right) \right]$ and $c(x) = c \left[ 1 + \eta_c \mathrm{sech}\left(\frac{x}{d}\right) \right]$ and $\eta_{\pm/c}$ are  dimensionless parameters determining the amount of $T_c$ enhancement. $d$ is the width of the effective interface.  A similar model was adopted previously to model the 3-K phase \cite{Maeno1998,Ando.99,Mao.01,Wang.17} of Sr$_2$RuO$_4$ \cite{Sigrist2001,Sigrist2019}. The instability condition is obtained by minimizing the free energy.  At the S/N interface we apply the De Gennes's boundary conditions $D_x \Phi_\nu|_{x=0+} = \Phi_{\nu}(x=0)/{\ell}$ \cite{DG1964}, where the extrapolation length $\ell$ is a characteristic length scale associated with the induced superconducting correlations. In contrast to conventional superconductors, here $\ell$  cannot be identified with  the superconducting correlation length of the DN segment, because it does not account for the physics of odd-frequency pairs \cite{TanakaGolubov2007,TSN_Review,Lindner.19}.  We therefore treat $\ell$ as a phenomenological parameter.  
	
	Figure \ref{Fig:T1}(b) illustrates the $H$ vs.\ $T$ phase diagram for the appearance of onset order at the interface for various {$d$} values. 
	The onset temperature and magnetic field are obtained by minimizing the free energy, Eq. \eqref{Eq:FE}, with De Gennes's boundary conditions with extrapolation $\ell=\xi_0$, where $\xi_0$ is the coherence length at $T=0$ for the bulk superconductor, see also S1 of\cite{SupMat}.  Within the GL formalism, $\To$ exceeds  {$T_c^{\rm bulk}$} in the low-field limit, and the $\To$ decreases with increasing  {field}. In our calculations, we find that the onset magnetic field in the low-$T$ limit is comparable to  $\tilde{H}^{\rm bulk}_{c2}\equiv \Phi_0/(2\pi \xi_0^2)$, which is a magnetic field scale of the order of orbital upper critical fields. However, quantitatively the onset magnetic field is much smaller compared to the experiments, despite a reasonable  $\To$ obtained from  theoretical calculations.
	
	In the low $T$ limit, the onset magnetic field exceeds the upper critical field of the bulk superconductor by a factor $\sim$\,100 \cite{CoSi2SC_b} and that of CoSi$_2$/Si films by $\sim$\,20 \cite{Chiu.21}.  The experimental onset magnetic field is above 2\,T and this value is comparable to the Pauli-limited field for the bulk superconductor ($\sim$\,2.4 T). The GL theory does not include the effect of Pauli paramagnetism or possible finite momentum pairing, suggesting that this could be the origin of the quantitative disagreement between the theory and experiment. The interface induced order survives up to 10--12 coherence lengths from the surface \cite{SupMat}. As shown in the Fig. \ref{Fig:T1}(b), $\To$ for the appearance of such order above  {$T_c^{\rm bulk}$} is very sensitive to the width of the interface. As the interface region becomes thinner, {$\To$} drops rapidly. In the devices with C49 phase, the interface region is expected to be relatively more disordered and its effective width is expected to be high compared to the {low-$\rho$} devices, due to smaller grain sizes or possible incomplete C49-C54 transformation (see S3 of  \onlinecite{SupMat}).
	
	Thus, Eq. \eqref{Eq:FE}  provides  an effective model for the CoSi$_2$/TiSi$_2$ heterojunctions. 
	The $\mathcal{F}_{me}$ term which suggest that the (CoSi$_2$/Si)/TiSi$_2$ system is an ideal system to explore {\itshape e.g.} the superconducting diode effect and charge transport effects \cite{Ando.20,Wakatsuki.17}. 
	Our results also demonstrate the stability of the APE, and  the prevalence of odd-frequency pairing in heterostructures
	\cite{TanakaGolubov2007,TSN_Review,Lindner.19}
	
	In conclusion, we have shown that the CoSi$_2$/TiSi$_2$ T-shaped proximity devices show strong evidence for two-component superconductivity. The tunneling spectra are robust against disorder, and the onset temperature for this two-component superconducting state exceeds the bulk $T_c$ with enhanced disorder. The large {SOC} which drives the two-component superconductivity is a result of the symmetry reduction  due to the CoSi$_2$/Si(100) interface, while the enhancement is driven by the CoSi$_2$/TiSi$_2$ interface.  We use quasi-classical theory to understand the tunneling spectra and a phenomenological theory to understand the enhanced superconductivity. The microscopic origin of this enhancement and the zero-bias conductance lineshape  is left for future studies.

	\begin{acknowledgments}
		\textit{Acknowledgments --} 
		We thank S. S. Yeh for experimental help. This work was supported by the Ministry of Science and Technology of Taiwan through grant numbers MOST 106-2112-M-009-007-MY4 and 110-2112-M-A49-015, and by the Ministry of Education of Taiwan through the Higher Education Sprout Project. VM, YL and FCZ are partially supported by NSFC grants 11674278 {and 11920101005}   and by the priority program of the Chinese Academy of Sciences grant No. XDB28000000, and  by the China Postdoctoral Science Foundation under grant No. 2020M670422 (YL).  
	\end{acknowledgments}
	
	%
	
	\clearpage

	\pagebreak 
	\newpage 
	
	\renewcommand*{\citenumfont}[1]{S#1}
	\renewcommand*{\bibnumfmt}[1]{[S#1]}
	\begin{widetext}
		\begin{center}
			\textbf{\large{}\textemdash{} Supplemental Material \textemdash{}}
			\par\end{center}{\large \par}
		\begin{center}
			\textbf{\large{}Enhanced two-component superconductivity in CoSi$_2$/TiSi$_2$ heterojunctions}
			\par\end{center}{\large \par}
		\begin{center}
			\textbf{Shao-Pin Chiu,$^{1,2,*}$ Vivek Mishra,$^{3,*}$ Yu Li,$^{3}$ Fu-Chun Zhang,$^{3,4,5}$ Stefan Kirchner,$^{1,2}$\\ and Juhn-Jong Lin$^{1,2}$}\\
			$^{1}$Department of Electrophysics, National Yang Ming Chiao Tung University, Hsinchu 30010, Taiwan\\
			$^{2}$Center for Emergent Functional Matter Science, NYCU, Hsinchu 30010, Taiwan\\
			$^{3}$Kavli Institute for Theoretical Sciences, University of Chinese Academy of Sciences, Beijing 100190, China\\
			$^{4}$CAS Center for Excellence in Topological Quantum Computation, University of Chinese Academy of Sciences, Beijing 100190, China\\
			$^5${{HKU-UCAS Joint Institute of Theoretical and Computational Physics at Beijing, University of Chinese Academy of Sciences, Beijing 100190, China}}
		\end{center}
		
		\vspace*{2mm}
		\begin{description}
			\item [{Summary}] 	Below we provide additional technical details, further experimental data and auxiliary numerical results supplementing the conclusions from the main text.
		\end{description}
		
	\end{widetext}

	\beginsupplement 
	
	\section{Ginzburg-Landau analysis of the enhanced interface two-component superconductivity}
\begin{figure}
		\includegraphics[width=0.86\linewidth]{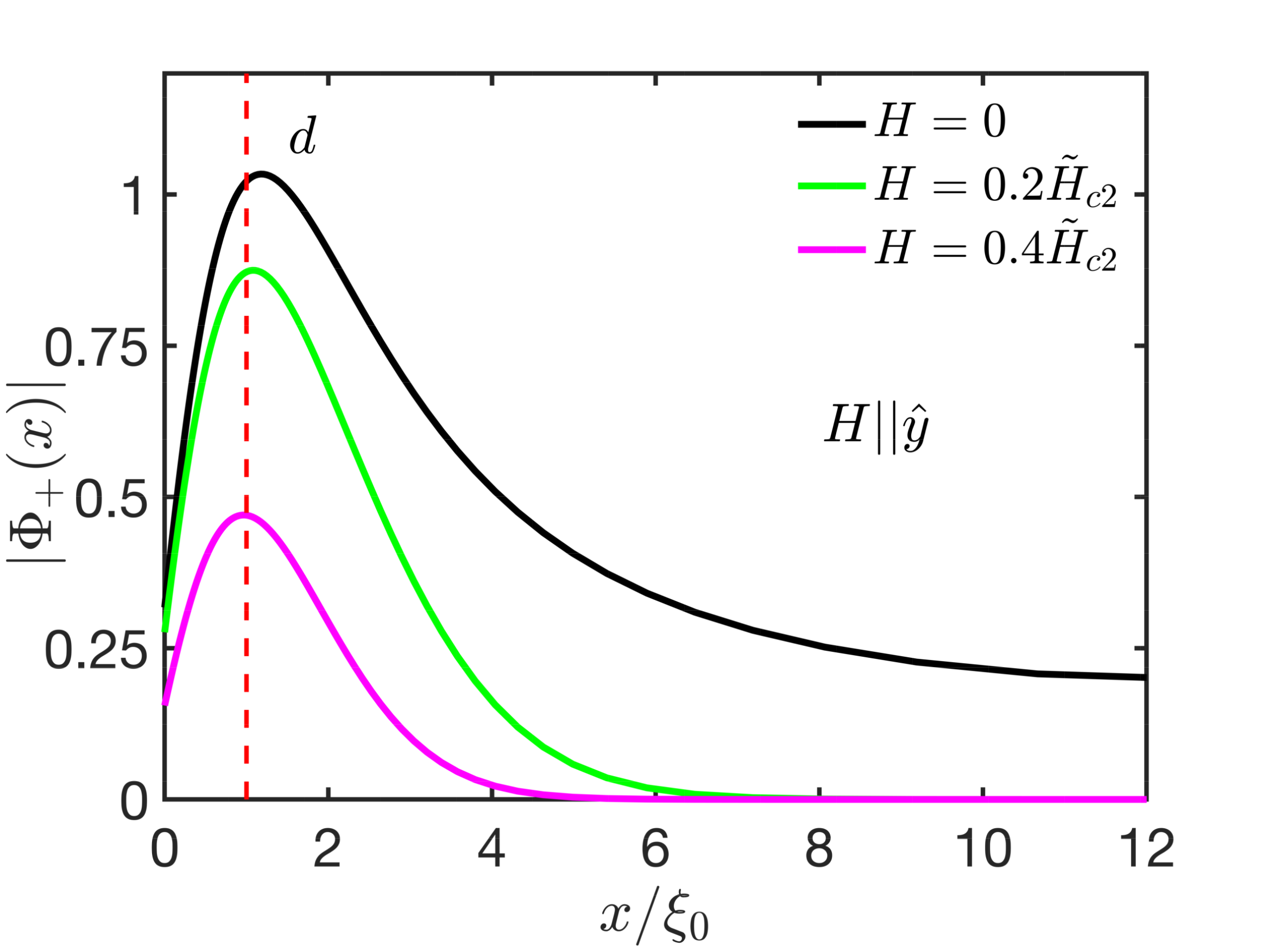}
		\caption{Absolute value of the dominant component $\Phi_+$ is plotted as a function of distance $x$ for several values of the magnetic field at $T=1.2 T_c$ for $d=\xi_0$ and $\ell=0.25\xi_0$. The S/N interface is located at $x=0$. The effective thickness of the interface layer {($d$)} is indicated.  }
		\label{Fig:ST1}
	\end{figure}

	To find the ground state, we minimize the Ginzburg-Landau (GL) free energy, which results in,
	
	\begin{eqnarray}
		K_s \partial_{x^2} \Phi_+ &=&  \left( a_+ + K_s \gamma^2 H^2 x^2 +  Q H^2 \right) \Phi_+ \nonumber \\
		&+&    2 b_+|\Phi_+|^2 \Phi_+  + c \Phi_{-} + 2 i K_{me+} H \partial_x \Phi_+,   \label{Eq:Ins1}
	\end{eqnarray}
	\begin{eqnarray}
		K_s \partial_{x^2} \Phi_- &=&  \left( a_- + K_s \gamma^2 H^2 x^2 +  Q H^2  \right) \Phi_- \nonumber \\
		&+&    2 b_-|\Phi_-|^2 \Phi_+  + c \Phi_{+} + 2 i K_{me-} H \partial_x \Phi_-. 
		\label{Eq:Ins2}
	\end{eqnarray}
	Here $\gamma=2\pi/\Phi_0$, where $\Phi_0$ is the magnetic flux quanta, and  we set  the vector potential $\vec{A}=(0,0,-Hx)$, which gives the magnetic field in the $\hat{y}$ direction. All lengths can be expressed in terms of the coherence length $\xi_0^{-2}=N_0 T_c /2 K_s$,  where the coefficient $K_s$ is $7\zeta(3) N_0 v_F^2 /64\pi^2  T_c^2$, $v_F$ is the average Fermi velocity, and $N_0$ is the density of states. Here $T_c$ refers to the bulk transition temperature.  We assume an isotropic Fermi-surface and ignore the difference in the Fermi velocities and density of states of two helical bands formed due to the spin-orbit coupling. Similarly, the magnetic field can be measured in unit of $\tilde{H}^{\rm bulk}_{c2}$, which is $\Phi_0/(2\pi \xi_0^2)$ and all energies in these equations can be expressed in units of $T_c$. The coefficient $a_\nu$ is {$N_0  (T-T_{c\nu})/2$}. The coefficient $Q$ is,
	\begin{equation}
		Q=\frac{7 \zeta(3) \mu_B^2 N_0 }{16 \pi^2 T_c^2},
	\end{equation}
	where $\mu_B$ is the Bohr magneton. The magneto-electric coefficient $K_{me\pm}$ has opposite sign for two helical bands \cite{SSigrist_Review}, and it reads,
	\begin{eqnarray}
		K_{me\pm} &= \mp \mu_B \frac{7\zeta(3)}{4\pi^2 T_c^2} \frac{N_0 v_F}{4}.
	\end{eqnarray}
	We further assume $b_\pm =b$. Equations \eqref{Eq:Ins1} and \eqref{Eq:Ins2} are solved with De Gennes's boundary condition at the superconductor and normal interface.   Figure \ref{Fig:ST1} shows the solution  for various values of the magnetic field, and it shows that the order parameters remain finite up to several coherence lengths from the {S/N} interface.
	
	\section{Additional Experimental Data}
	
\begin{figure}[t!]
		\centering
		\includegraphics[width= 0.36\textwidth]{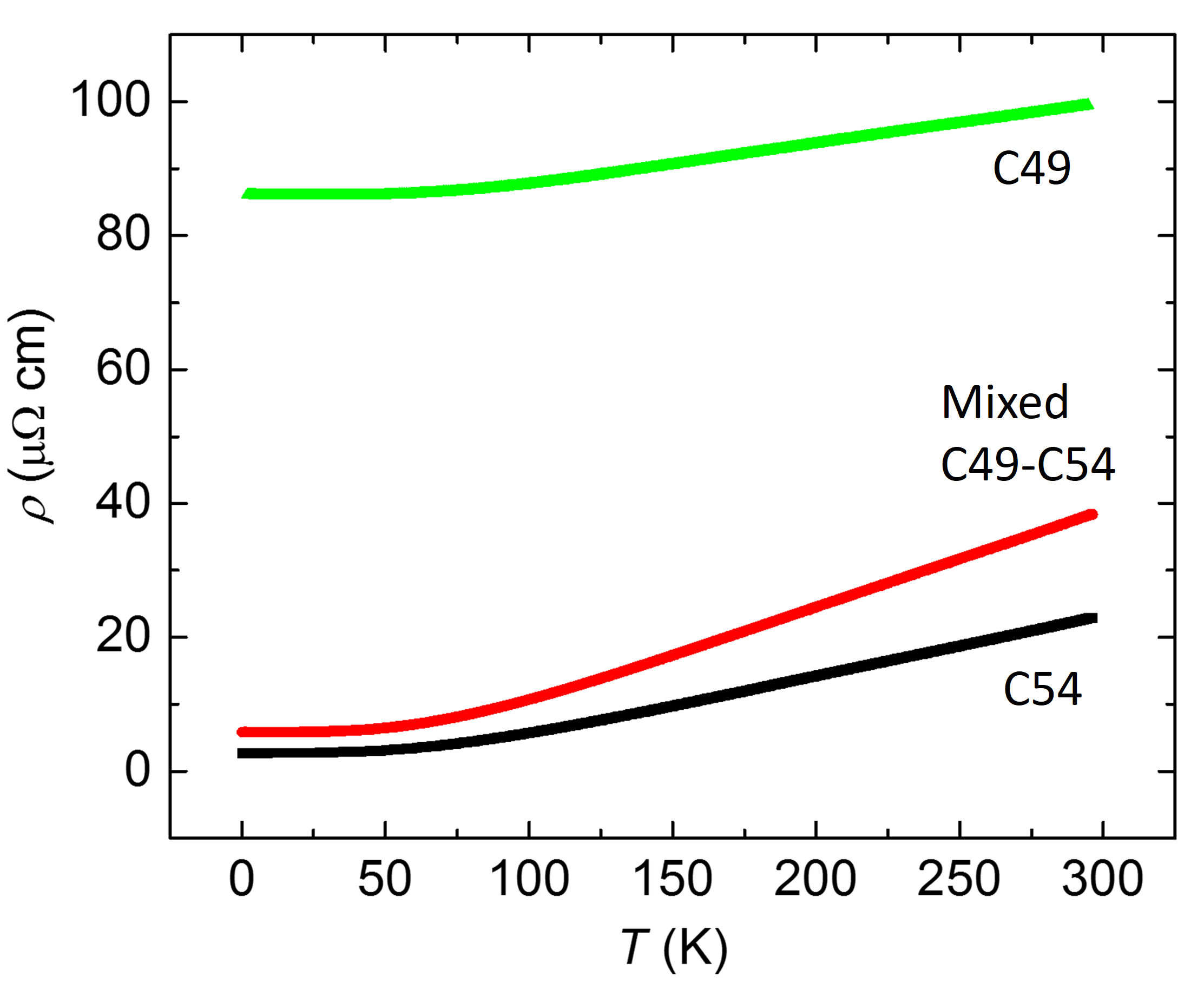} 
		\caption{Resistivity as a function of temperature for a C49-, a mixed C49/C54-, and a C54-phased TiSi$_2$/Si(100) films, as indicated. The films are all 125-nm thick. The $\rho (T)$ curve in each case reveals typical electrical-transport behavior of a weakly disordered metal, {\itshape i.e.}, $\rho$ decreases with decreasing $T$. 
		}
		\label{figS2}
	\end{figure}

\begin{figure}[t!]
		\centering
		\includegraphics[width= 0.4\textwidth]{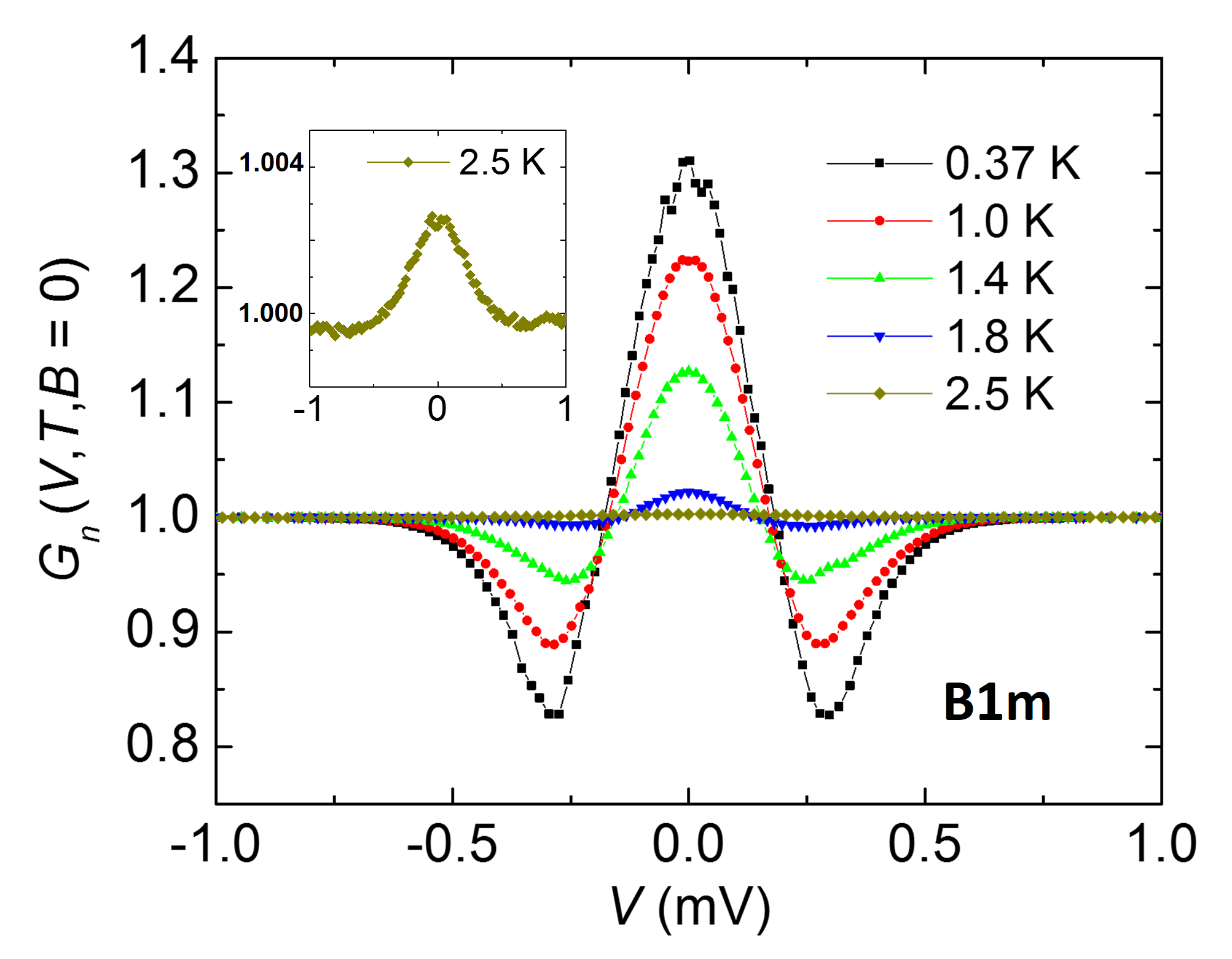} 
		\caption{Finite-bias $G_n(V,T,0)$ at several $T$ values of device B1m. The amplitude of ZBCP gradually decreases with increasing $T$. The inset indicates that the ZBCP persists up to at least 2.5\,K.
		}
		\label{figS3}
	\end{figure}

\begin{figure}[t!]
		\centering
		\includegraphics[width= 0.48\textwidth]{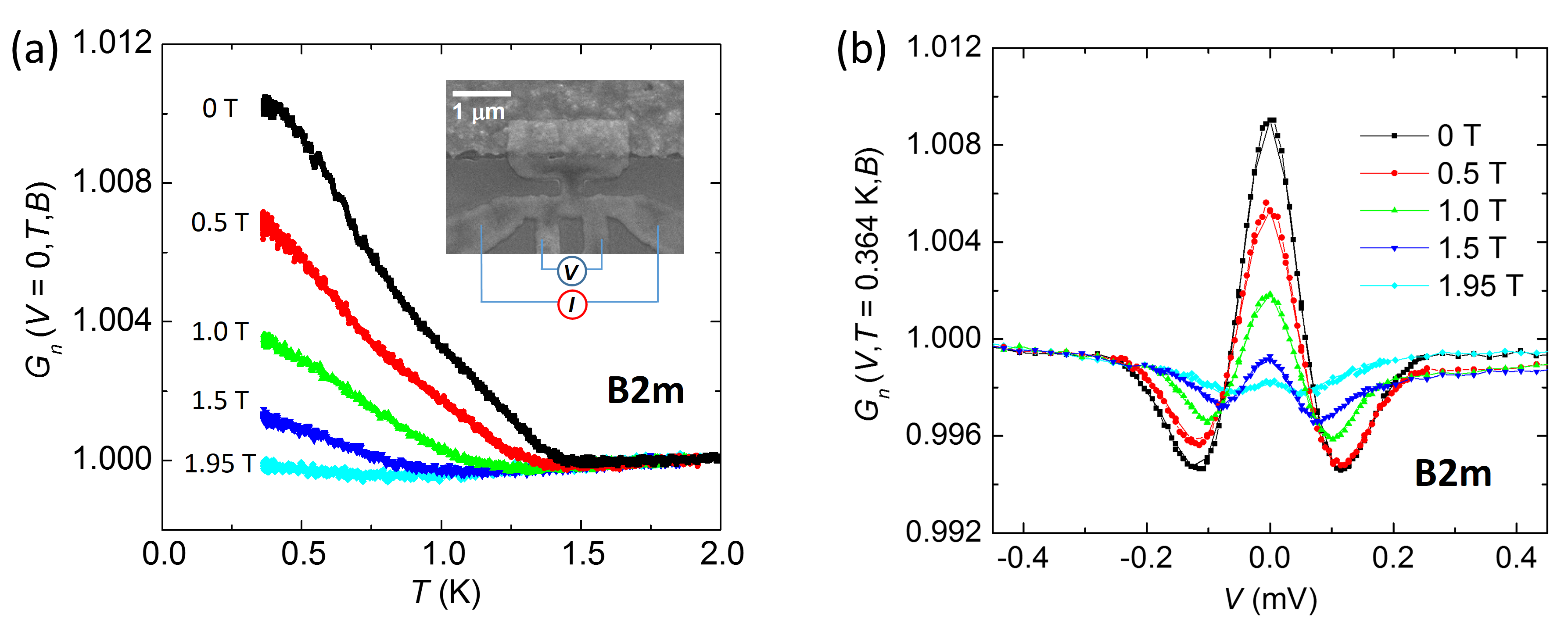} 
		\caption{Normalized conductance spectra of device B2m. (a) Zero-bias $G_n(0,T,B)$ in several $B$ fields. With increasing $B$, $G_n$ is gradually suppressed. The inset shows an SEM image of the device, with a schematic 4-probe configuration. (b) Finite-bias $G_n(V,0.364\,{\rm K},B)$ in several $B$ fields. A close inspection indicates that the ZBCP persists up to at least 1.95\,T, even though the onset temperature ($T_c^{\rm onset} = 1.42$\,K) is not enhanced in this device.
		}
		\label{figS4}
	\end{figure}

\begin{figure}[t!]
		\centering
		\includegraphics[width= 0.48\textwidth]{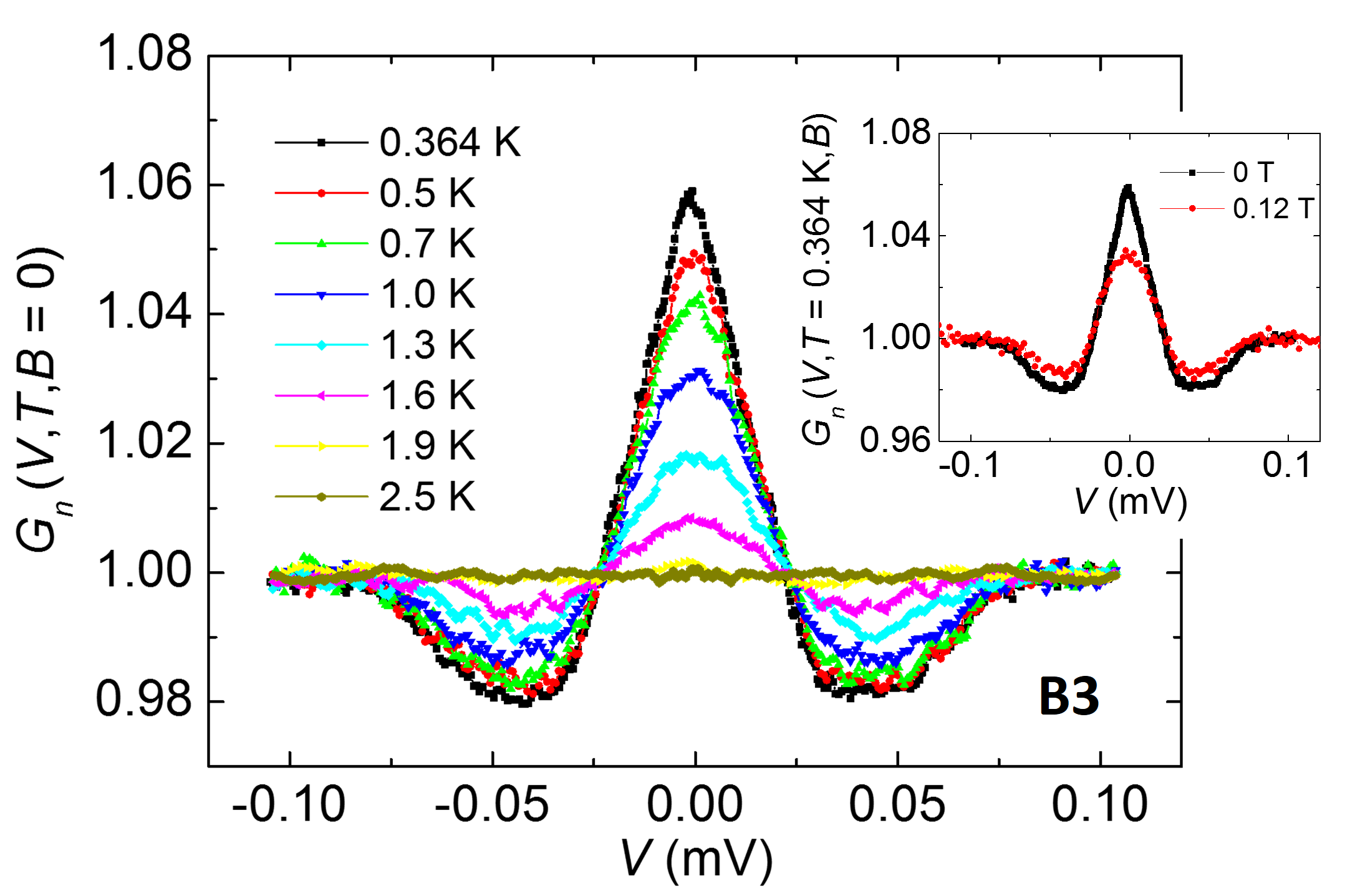} 
		\caption{Finite-bias $G_n(V,T,0)$ for device B3 at several $T$ values. The amplitude of ZBCP gradually decreases with increasing $T$. The inset depicts that the ZBCP is only weakly suppressed in $B=0.12$\,T and at $T=0.364$\,K. 
		}
		\label{figS5}
	\end{figure}

\begin{figure}[t!]
		\centering
		\includegraphics[width= 0.48\textwidth]{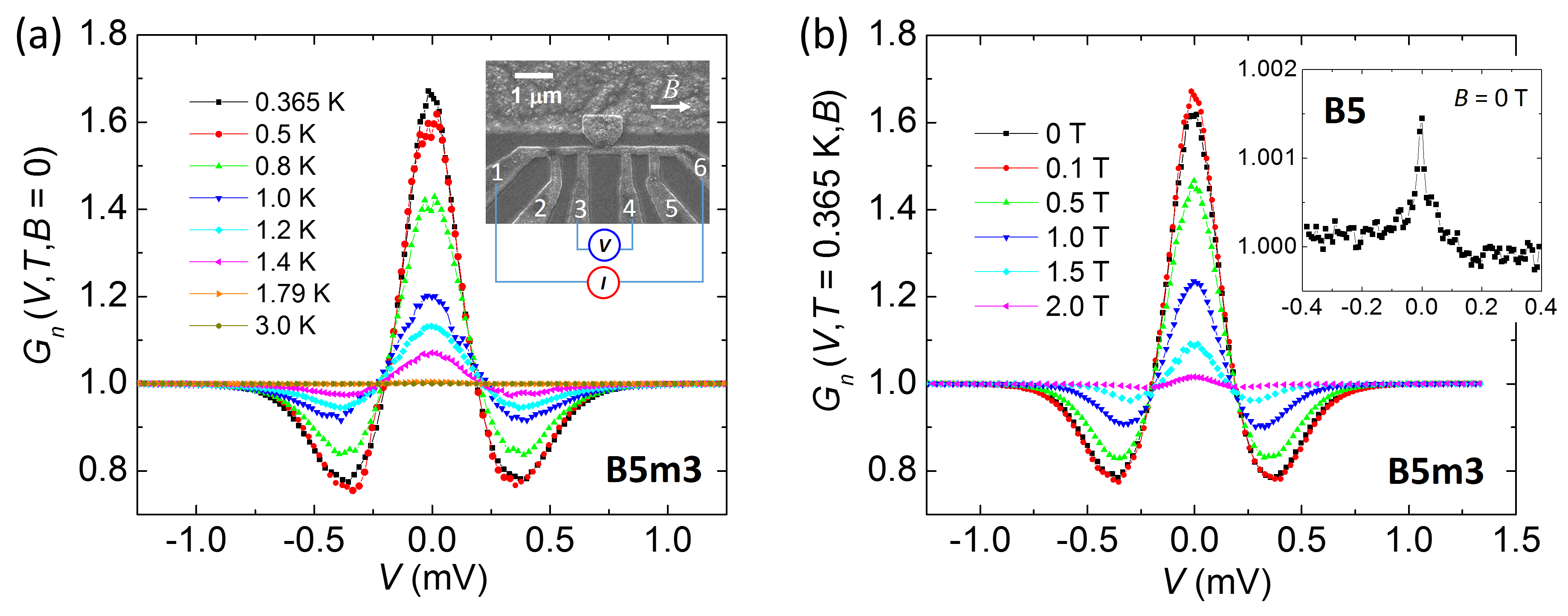} 
		\caption{Normalized conductance spectra of device B5m3 measured with the VE pair (3,4). (a) Finite-bias $G_n(V,T,0)$ at several $T$ values. The amplitude of ZBCP gradually decreases with increasing $T$, persisting up to about 1.79\,K. The inset shows an SEM image of the device, with a schematic 4-probe configuration. (b) Finite-bias $G_n(V,0.365\,{\rm K},B)$ in several $B$ fields. The amplitude of ZBCP is gradually suppressed by increasing $B$. The inset shows that the $G_n(V,0.37\,{\rm K},0)$ of the as-grown device B5, also measured with the VE pair (3,4), is significantly smaller than that of device B5m3.  
		}
		\label{figS6}
	\end{figure}

	Auxiliary experimental data on the electrical-transport properties of normal-metal TiSi$_2$ films (Fig. \ref{figS2}), as well as additional conductance spectra of CoSi$_2$/TiSi$_2$ {T}-shaped superconducting proximity structures (Figs. \ref{figS3} to \ref{figS6}) are presented. A brief discussion of the data is provided in each of the figure captions.
	
	\newpage
	\section{Device Characterizations} 
	This supplemental section provides a compilation of the device parameters for the CoSi$_2$/TiSi$_2$ {T}-shaped proximity structures discussed in the main text.
	
	\textbf{Formation of C49 and C54 TiSi$_2$ phases.}
	Titanium films were deposited on {an electron-beam lithographically patterned} Si(100) substrate, and {subsequently} thermally annealed to form {the normal-metal} TiSi$_2$ {component in the {T}-shaped proximity structure}. The phase formation sequence of titanium silicide followed the reaction path as the thermal annealing temperature was raised: Ti/Si $\rightarrow$ amorphous TiSi$_x$ $\rightarrow$ polycrystalline C49 TiSi$_2$ $\rightarrow$ polycrystalline C54 TiSi$_2$ \cite{SMa.91,SChen.04}. The C49 (C54) TiSi$_2$ phase was obtained by annealing at 750$^\circ$C (800$^\circ$C) for about 1 h \cite{SChiu.21}. A mixed C49-C54 TiSi$_2$ phase could also be obtained by applying an intermediate annealing temperature 780$^\circ$C. Cross-sectional transmission electron microscopy studies revealed that the C49 phase had a smaller grain size ($\sim$ 50--300 nm), compared with that ($\sim$ 300--1000 nm) of the C54 phase. Moreover, the carrier concentration in the C49 phase ($\approx$ 1.5$\times$$10^{22}$ cm$^{-3}$) is more than one order of magnitude lower than that ($\approx$ 3$\times$$10^{23}$ cm$^{-3}$) in the C54 phase \cite{SMammoliti.02}. Thus, $\rho$(C49) $\gg$ $\rho$(C54).
	
	\textbf{Thermal cycling effect and device labeling.}
	After the conductance spectra of the as-grown device B1 (B2) were measured at low temperatures, the device was warmed up from 0.36\,K to 300\,K and cooled down again for a second-run measurement of the conductance spectra. The device was then labeled B1m (B2m), because the device properties (e.g., the transparency of the S/N interface as well as the defect configurations of the TiSi$_2$ component) were often sensitive to thermal cycling and repeated finite-bias measurements. Device B5 was intentionally subject to three times of thermal cycling to check the robustness and reproducibility of the conductance spectra. Thus, B5m3 denotes the fourth-run measurement of device B5. 
	
	In the course of this study, we find that thermal cycling from liquid-helium temperature to 300\,K could result in modifications of the CoSi$_2$/TiSi$_2$ interface and/or rearrangements of defect configurations in the TiSi$_2$ component. We found that thermal cycling often caused a decrease in $\rho$(TiSi$_2$), but occasionally it resulted in an increase in $\rho$(TiSi$_2$). Moreover, we have observed that an enhanced $T_c^{\mbox{\tiny onset}}$ value is more frequently found in those T-shaped structures made of high-$\rho$ TiSi$_2$ component. In other words, in those T-shaped structures whose $\rho$(TiSi$_2$) values were reduced by thermal cycling, their $T_c^{\mbox{\tiny onset}}$ values were also reduced.

	\begin{widetext}
		
		\renewcommand{\arraystretch}{1.2}
		\begin{table*}[h!]
			\centering
			\begin{tabular}{|l|cccccccccc|}
				\toprule
				\rowcolor{gray!50}
				{\rm Device~~~}%
				&\shortstack{$T_{\rm a}$\\($^\circ$C)}%
				&\shortstack{$L_1$\\($\mu$m)}%
				&\shortstack{$L_2$\\($\mu$m)}%
				&\shortstack{$R_N$\\($\Omega$)}%
				&\shortstack{$\rho_N$\\($\mu \Omega$ cm)}%
				&\shortstack{$\Delta G_n$\\(at 0.37\,K)}%
				&\shortstack{FWHM\\(meV)}%
				&\shortstack{$E_{\rm Th}$\\(meV)}%
				&\shortstack{$D$\\(cm$^2$/s)}%
				&\shortstack{$T_c^{\mbox{\tiny onset}}$\\(K)}
				\\
				\midrule \midrule
				{\rm B1}  & 750 & 0.42 & 0.39  & 57.5 & 197 & 112\% & 0.18  & 0.0012 & 3.1  &  2.94\\
				\midrule
				{\rm B1m} & 750 & 0.42 & 0.39 & 55.8 & 191  & 31\% & 0.22  & 0.0012 & 3.2 & 2.84\\
				\midrule
				{\rm B2} & 750 & 0.20 & 0.66 & 12.0 & 79.8  & 9.3\% & 0.04  & 0.013 & 7.7 &2.33 \\
				\midrule 
				{\rm B2m}  & 750  & 0.20 & 0.66  & 12.7 & 84.3  & 1.0\%   &0.056   &0.012 & 7.3 & 1.42\\
				\midrule
				{\rm B3}  & 780  & 0.23 & 0.10 & 0.346 & 2.14  & 5.9\% & 0.026    &0.13 & 106 & 2.14\\
				\midrule
				{\rm B5}  & 800  & 0.45 & 0.32  & 1.31 & 3.34 & 0.13\% & 0.03   & 0.023 & 67.9 & 1.50 \\
				\midrule
				{\rm B5m3}  & 800  & 0.45 & 0.32  & 19.1 & 48.6 & 66\%   & 0.20  & 0.004 & 12.7 & 2.10\\
				\bottomrule
			\end{tabular}
			\caption{\label{table_1}%
				\textbf{Device parameters of CoSi$_2$/TiSi$_2$ T-shaped superconducting proximity structures.} 
				$T_{\rm a}$ is the thermal annealing temperature for the formation of the TiSi$_2$ component. $L_1$ and $L_2$ are defined in Fig. 1(a) in the main text. $R_N$ ($\rho_N$) is the residual resistance (resistivity) of the TiSi$_2$ component at 4 K. $\Delta G_n = G_n - 1$ is the increase in normalized differential conductance above the normal-state value (=\,1). FWHM is the full-width at half-maximum of the ZBCP. Thouless energy is defined by $E_{\rm Th} \approx \hbar D/L_1^2$, where $D$ is the electron diffusion constant of the TiSi$_2$ component. $T^{\mbox{\tiny onset}}_c$ is the onset temperature of the  APE, defined by $\Delta G_n(T = T_c^{\mbox{\tiny onset}}) = 10^{-3} \times \Delta G_n(T = 0.37\,{\rm K})$.
			}\label{ST1}
		\end{table*}
	\end{widetext}
	
	%


\begin{thebibliography}{46}%
		\makeatletter
		\providecommand \@ifxundefined [1]{%
			\@ifx{#1\undefined}
		}%
		\providecommand \@ifnum [1]{%
			\ifnum #1\expandafter \@firstoftwo
			\else \expandafter \@secondoftwo
			\fi
		}%
		\providecommand \@ifx [1]{%
			\ifx #1\expandafter \@firstoftwo
			\else \expandafter \@secondoftwo
			\fi
		}%
		\providecommand \natexlab [1]{#1}%
		\providecommand \enquote  [1]{``#1''}%
		\providecommand \bibnamefont  [1]{#1}%
		\providecommand \bibfnamefont [1]{#1}%
		\providecommand \citenamefont [1]{#1}%
		\providecommand \href@noop [0]{\@secondoftwo}%
		\providecommand \href [0]{\begingroup \@sanitize@url \@href}%
		\providecommand \@href[1]{\@@startlink{#1}\@@href}%
		\providecommand \@@href[1]{\endgroup#1\@@endlink}%
		\providecommand \@sanitize@url [0]{\catcode `\\12\catcode `\$12\catcode
			`\&12\catcode `\#12\catcode `\^12\catcode `\_12\catcode `\%12\relax}%
		\providecommand \@@startlink[1]{}%
		\providecommand \@@endlink[0]{}%
		\providecommand \url  [0]{\begingroup\@sanitize@url \@url }%
		\providecommand \@url [1]{\endgroup\@href {#1}{\urlprefix }}%
		\providecommand \urlprefix  [0]{URL }%
		\providecommand \Eprint [0]{\href }%
		\providecommand \doibase [0]{http://dx.doi.org/}%
		\providecommand \selectlanguage [0]{\@gobble}%
		\providecommand \bibinfo  [0]{\@secondoftwo}%
		\providecommand \bibfield  [0]{\@secondoftwo}%
		\providecommand \translation [1]{[#1]}%
		\providecommand \BibitemOpen [0]{}%
		\providecommand \bibitemStop [0]{}%
		\providecommand \bibitemNoStop [0]{.\EOS\space}%
		\providecommand \EOS [0]{\spacefactor3000\relax}%
		\providecommand \BibitemShut  [1]{\csname bibitem#1\endcsname}%
		\let\auto@bib@innerbib\@empty
		\bibitem [{\citenamefont {Read}\ and\ \citenamefont {Green}(2000)}]{Read.00}%
		\BibitemOpen
		\bibfield  {author} {\bibinfo {author} {\bibfnamefont {N.}~\bibnamefont
				{Read}}\ and\ \bibinfo {author} {\bibfnamefont {D.}~\bibnamefont {Green}},\
		}\href {\doibase 10.1103/PhysRevB.61.10267} {\bibfield  {journal} {\bibinfo
				{journal} {Phys. Rev. B}\ }\textbf {\bibinfo {volume} {61}},\ \bibinfo
			{pages} {10267} (\bibinfo {year} {2000})}\BibitemShut {NoStop}%
		\bibitem [{\citenamefont {Nayak}\ \emph {et~al.}(2008)\citenamefont {Nayak},
			\citenamefont {Simon}, \citenamefont {Stern}, \citenamefont {Freedman},\ and\
			\citenamefont {Das~Sarma}}]{Nayak.08}%
		\BibitemOpen
		\bibfield  {author} {\bibinfo {author} {\bibfnamefont {C.}~\bibnamefont
				{Nayak}}, \bibinfo {author} {\bibfnamefont {S.~H.}\ \bibnamefont {Simon}},
			\bibinfo {author} {\bibfnamefont {A.}~\bibnamefont {Stern}}, \bibinfo
			{author} {\bibfnamefont {M.}~\bibnamefont {Freedman}}, \ and\ \bibinfo
			{author} {\bibfnamefont {S.}~\bibnamefont {Das~Sarma}},\ }\href {\doibase
			10.1103/RevModPhys.80.1083} {\bibfield  {journal} {\bibinfo  {journal} {Rev.
					Mod. Phys.}\ }\textbf {\bibinfo {volume} {80}},\ \bibinfo {pages} {1083}
			(\bibinfo {year} {2008})}\BibitemShut {NoStop}%
		\bibitem [{\citenamefont {Qi}\ and\ \citenamefont {Zhang}(2011)}]{Qi.11}%
		\BibitemOpen
		\bibfield  {author} {\bibinfo {author} {\bibfnamefont {X.-L.}\ \bibnamefont
				{Qi}}\ and\ \bibinfo {author} {\bibfnamefont {S.-C.}\ \bibnamefont {Zhang}},\
		}\href {\doibase 10.1103/RevModPhys.83.1057} {\bibfield  {journal} {\bibinfo
				{journal} {Rev. Mod. Phys.}\ }\textbf {\bibinfo {volume} {83}},\ \bibinfo
			{pages} {1057} (\bibinfo {year} {2011})}\BibitemShut {NoStop}%
		\bibitem [{\citenamefont {Alicea}(2012)}]{Alicea.12}%
		\BibitemOpen
		\bibfield  {author} {\bibinfo {author} {\bibfnamefont {J.}~\bibnamefont
				{Alicea}},\ }\href {\doibase 10.1088/0034-4885/75/7/076501} {\bibfield
			{journal} {\bibinfo  {journal} {Reports on Progress in Physics}\ }\textbf
			{\bibinfo {volume} {75}},\ \bibinfo {pages} {076501} (\bibinfo {year}
			{2012})}\BibitemShut {NoStop}%
		\bibitem [{\citenamefont {Asano}\ \emph {et~al.}(2007)\citenamefont {Asano},
			\citenamefont {Tanaka}, \citenamefont {Golubov},\ and\ \citenamefont
			{Kashiwaya}}]{Asano2007}%
		\BibitemOpen
		\bibfield  {author} {\bibinfo {author} {\bibfnamefont {Y.}~\bibnamefont
				{Asano}}, \bibinfo {author} {\bibfnamefont {Y.}~\bibnamefont {Tanaka}},
			\bibinfo {author} {\bibfnamefont {A.~A.}\ \bibnamefont {Golubov}}, \ and\
			\bibinfo {author} {\bibfnamefont {S.}~\bibnamefont {Kashiwaya}},\ }\href
		{\doibase 10.1103/PhysRevLett.99.067005} {\bibfield  {journal} {\bibinfo
				{journal} {Phys. Rev. Lett.}\ }\textbf {\bibinfo {volume} {99}},\ \bibinfo
			{pages} {067005} (\bibinfo {year} {2007})}\BibitemShut {NoStop}%
		\bibitem [{\citenamefont {Tanaka}\ \emph {et~al.}(2005)\citenamefont {Tanaka},
			\citenamefont {Asano}, \citenamefont {Golubov},\ and\ \citenamefont
			{Kashiwaya}}]{Tanaka2005}%
		\BibitemOpen
		\bibfield  {author} {\bibinfo {author} {\bibfnamefont {Y.}~\bibnamefont
				{Tanaka}}, \bibinfo {author} {\bibfnamefont {Y.}~\bibnamefont {Asano}},
			\bibinfo {author} {\bibfnamefont {A.~A.}\ \bibnamefont {Golubov}}, \ and\
			\bibinfo {author} {\bibfnamefont {S.}~\bibnamefont {Kashiwaya}},\ }\href
		{\doibase 10.1103/PhysRevB.72.140503} {\bibfield  {journal} {\bibinfo
				{journal} {Phys. Rev. B}\ }\textbf {\bibinfo {volume} {72}},\ \bibinfo
			{pages} {140503} (\bibinfo {year} {2005})}\BibitemShut {NoStop}%
		\bibitem [{\citenamefont {Tanaka}\ \emph {et~al.}(2006)\citenamefont {Tanaka},
			\citenamefont {Asano}, \citenamefont {Golubov},\ and\ \citenamefont
			{Kashiwaya}}]{Tanaka2005_ERR}%
		\BibitemOpen
		\bibfield  {author} {\bibinfo {author} {\bibfnamefont {Y.}~\bibnamefont
				{Tanaka}}, \bibinfo {author} {\bibfnamefont {Y.}~\bibnamefont {Asano}},
			\bibinfo {author} {\bibfnamefont {A.~A.}\ \bibnamefont {Golubov}}, \ and\
			\bibinfo {author} {\bibfnamefont {S.}~\bibnamefont {Kashiwaya}},\ }\href
		{\doibase 10.1103/PhysRevB.73.059901} {\bibfield  {journal} {\bibinfo
				{journal} {Phys. Rev. B}\ }\textbf {\bibinfo {volume} {73}},\ \bibinfo
			{pages} {059901} (\bibinfo {year} {2006})}\BibitemShut {NoStop}%
		\bibitem [{\citenamefont {Tanaka}\ and\ \citenamefont
			{Golubov}(2007)}]{TanakaGolubov2007}%
		\BibitemOpen
		\bibfield  {author} {\bibinfo {author} {\bibfnamefont {Y.}~\bibnamefont
				{Tanaka}}\ and\ \bibinfo {author} {\bibfnamefont {A.~A.}\ \bibnamefont
				{Golubov}},\ }\href {\doibase 10.1103/PhysRevLett.98.037003} {\bibfield
			{journal} {\bibinfo  {journal} {Phys. Rev. Lett.}\ }\textbf {\bibinfo
				{volume} {98}},\ \bibinfo {pages} {037003} (\bibinfo {year}
			{2007})}\BibitemShut {NoStop}%
		\bibitem [{\citenamefont {Courtois}\ \emph {et~al.}(1999)\citenamefont
			{Courtois}, \citenamefont {Charlat}, \citenamefont {Gandit}, \citenamefont
			{Mailly},\ and\ \citenamefont {Pannetier}}]{Courtois1999}%
		\BibitemOpen
		\bibfield  {author} {\bibinfo {author} {\bibfnamefont {H.}~\bibnamefont
				{Courtois}}, \bibinfo {author} {\bibfnamefont {P.}~\bibnamefont {Charlat}},
			\bibinfo {author} {\bibfnamefont {P.}~\bibnamefont {Gandit}}, \bibinfo
			{author} {\bibfnamefont {D.}~\bibnamefont {Mailly}}, \ and\ \bibinfo {author}
			{\bibfnamefont {B.}~\bibnamefont {Pannetier}},\ }\href {\doibase
			10.1023/A:1021885617107} {\bibfield  {journal} {\bibinfo  {journal} {Journal
					of Low Temperature Physics}\ }\textbf {\bibinfo {volume} {116}},\ \bibinfo
			{pages} {187} (\bibinfo {year} {1999})}\BibitemShut {NoStop}%
		\bibitem [{\citenamefont {Mackenzie}\ \emph {et~al.}(2017)\citenamefont
			{Mackenzie}, \citenamefont {Scaffidi}, \citenamefont {Hicks},\ and\
			\citenamefont {Maeno}}]{Mackenzie2017}%
		\BibitemOpen
		\bibfield  {author} {\bibinfo {author} {\bibfnamefont {A.~P.}\ \bibnamefont
				{Mackenzie}}, \bibinfo {author} {\bibfnamefont {T.}~\bibnamefont {Scaffidi}},
			\bibinfo {author} {\bibfnamefont {C.~W.}\ \bibnamefont {Hicks}}, \ and\
			\bibinfo {author} {\bibfnamefont {Y.}~\bibnamefont {Maeno}},\ }\href
		{\doibase 10.1038/s41535-017-0045-4} {\bibfield  {journal} {\bibinfo
				{journal} {npj Quantum Materials}\ }\textbf {\bibinfo {volume} {2}},\
			\bibinfo {pages} {40} (\bibinfo {year} {2017})}\BibitemShut {NoStop}%
		\bibitem [{\citenamefont {{Pereiro}}\ \emph {et~al.}(2011)\citenamefont
			{{Pereiro}}, \citenamefont {{Petrovic}}, \citenamefont {{Panagopoulos}},\
			and\ \citenamefont {{Bo{\v{z}}ovi{\'c}}}}]{Bozovic.11}%
		\BibitemOpen
		\bibfield  {author} {\bibinfo {author} {\bibfnamefont {J.}~\bibnamefont
				{{Pereiro}}}, \bibinfo {author} {\bibfnamefont {A.}~\bibnamefont
				{{Petrovic}}}, \bibinfo {author} {\bibfnamefont {C.}~\bibnamefont
				{{Panagopoulos}}}, \ and\ \bibinfo {author} {\bibfnamefont {I.}~\bibnamefont
				{{Bo{\v{z}}ovi{\'c}}}},\ }\href@noop {} {\enquote {\bibinfo {title}
				{{Interface superconductivity: History, development and prospects}},}\ }
		(\bibinfo {year} {2011}),\ \Eprint {http://arxiv.org/abs/1111.4194}
		{arXiv:1111.4194} \BibitemShut {NoStop}%
		\bibitem [{\citenamefont {Chiu}\ \emph
			{et~al.}(2021{\natexlab{a}})\citenamefont {Chiu}, \citenamefont {Tsuei},
			\citenamefont {Yeh}, \citenamefont {Zhang}, \citenamefont {Kirchner},\ and\
			\citenamefont {Lin}}]{Chiu.21}%
		\BibitemOpen
		\bibfield  {author} {\bibinfo {author} {\bibfnamefont {S.-P.}\ \bibnamefont
				{Chiu}}, \bibinfo {author} {\bibfnamefont {C.~C.}\ \bibnamefont {Tsuei}},
			\bibinfo {author} {\bibfnamefont {S.-S.}\ \bibnamefont {Yeh}}, \bibinfo
			{author} {\bibfnamefont {F.-C.}\ \bibnamefont {Zhang}}, \bibinfo {author}
			{\bibfnamefont {S.}~\bibnamefont {Kirchner}}, \ and\ \bibinfo {author}
			{\bibfnamefont {J.-J.}\ \bibnamefont {Lin}},\ }\href {\doibase
			10.1126/sciadv.abg6569} {\bibfield  {journal} {\bibinfo  {journal} {Science
					Advances}\ }\textbf {\bibinfo {volume} {7}},\ \bibinfo {pages} {eabg6569}
			(\bibinfo {year} {2021}{\natexlab{a}})}\BibitemShut {NoStop}%
		\bibitem [{\citenamefont {Chiu}\ \emph
			{et~al.}(2021{\natexlab{b}})\citenamefont {Chiu}, \citenamefont {Lai},\ and\
			\citenamefont {Lin}}]{Chiu.21jjap}%
		\BibitemOpen
		\bibfield  {author} {\bibinfo {author} {\bibfnamefont {S.-P.}\ \bibnamefont
				{Chiu}}, \bibinfo {author} {\bibfnamefont {W.-L.}\ \bibnamefont {Lai}}, \
			and\ \bibinfo {author} {\bibfnamefont {J.-J.}\ \bibnamefont {Lin}},\ }\href
		{\doibase 10.35848/1347-4065/ac1693} {\bibfield  {journal} {\bibinfo
				{journal} {Japanese Journal of Applied Physics}\ }\textbf {\bibinfo {volume}
				{60}},\ \bibinfo {pages} {088002} (\bibinfo {year}
			{2021}{\natexlab{b}})}\BibitemShut {NoStop}%
		\bibitem [{\citenamefont {Matthias}(1952)}]{CoSi2SC}%
		\BibitemOpen
		\bibfield  {author} {\bibinfo {author} {\bibfnamefont {B.~T.}\ \bibnamefont
				{Matthias}},\ }\href {\doibase 10.1103/PhysRev.87.380} {\bibfield  {journal}
			{\bibinfo  {journal} {Phys. Rev.}\ }\textbf {\bibinfo {volume} {87}},\
			\bibinfo {pages} {380} (\bibinfo {year} {1952})}\BibitemShut {NoStop}%
		\bibitem [{\citenamefont {Matthias}\ and\ \citenamefont
			{Hulm}(1953)}]{CoSi2SC_b}%
		\BibitemOpen
		\bibfield  {author} {\bibinfo {author} {\bibfnamefont {B.~T.}\ \bibnamefont
				{Matthias}}\ and\ \bibinfo {author} {\bibfnamefont {J.~K.}\ \bibnamefont
				{Hulm}},\ }\href {\doibase 10.1103/PhysRev.89.439} {\bibfield  {journal}
			{\bibinfo  {journal} {Phys. Rev.}\ }\textbf {\bibinfo {volume} {89}},\
			\bibinfo {pages} {439} (\bibinfo {year} {1953})}\BibitemShut {NoStop}%
		\bibitem [{\citenamefont {Tsutsumi}\ \emph {et~al.}(1997)\citenamefont
			{Tsutsumi}, \citenamefont {Takayanagi},\ and\ \citenamefont
			{Hirano}}]{CoSi2SH}%
		\BibitemOpen
		\bibfield  {author} {\bibinfo {author} {\bibfnamefont {K.}~\bibnamefont
				{Tsutsumi}}, \bibinfo {author} {\bibfnamefont {S.}~\bibnamefont
				{Takayanagi}}, \ and\ \bibinfo {author} {\bibfnamefont {T.}~\bibnamefont
				{Hirano}},\ }\href {\doibase 10.1016/S0921-4526(97)00188-9} {\bibfield
			{journal} {\bibinfo  {journal} {Physica B: Condensed Matter}\ }\textbf
			{\bibinfo {volume} {237-238}},\ \bibinfo {pages} {310 } (\bibinfo {year}
			{1997})}\BibitemShut {NoStop}%
		\bibitem [{\citenamefont {Mattheiss}\ and\ \citenamefont
			{Hamann}(1988)}]{CoSi2ES}%
		\BibitemOpen
		\bibfield  {author} {\bibinfo {author} {\bibfnamefont {L.~F.}\ \bibnamefont
				{Mattheiss}}\ and\ \bibinfo {author} {\bibfnamefont {D.~R.}\ \bibnamefont
				{Hamann}},\ }\href {\doibase 10.1103/PhysRevB.37.10623} {\bibfield  {journal}
			{\bibinfo  {journal} {Phys. Rev. B}\ }\textbf {\bibinfo {volume} {37}},\
			\bibinfo {pages} {10623} (\bibinfo {year} {1988})}\BibitemShut {NoStop}%
		\bibitem [{\citenamefont {Mishra}\ \emph {et~al.}(2021)\citenamefont {Mishra},
			\citenamefont {Li}, \citenamefont {Zhang},\ and\ \citenamefont
			{Kirchner}}]{Mishra.21}%
		\BibitemOpen
		\bibfield  {author} {\bibinfo {author} {\bibfnamefont {V.}~\bibnamefont
				{Mishra}}, \bibinfo {author} {\bibfnamefont {Y.}~\bibnamefont {Li}}, \bibinfo
			{author} {\bibfnamefont {F.-C.}\ \bibnamefont {Zhang}}, \ and\ \bibinfo
			{author} {\bibfnamefont {S.}~\bibnamefont {Kirchner}},\ }\href {\doibase
			10.1103/PhysRevB.103.184505} {\bibfield  {journal} {\bibinfo  {journal}
				{Phys. Rev. B}\ }\textbf {\bibinfo {volume} {103}},\ \bibinfo {pages}
			{184505} (\bibinfo {year} {2021})}\BibitemShut {NoStop}%
		\bibitem [{\citenamefont {Chiu}\ \emph {et~al.}(2017)\citenamefont {Chiu},
			\citenamefont {Yeh}, \citenamefont {Chiou}, \citenamefont {Chou},
			\citenamefont {Lin},\ and\ \citenamefont {Tsuei}}]{Chiu.17}%
		\BibitemOpen
		\bibfield  {author} {\bibinfo {author} {\bibfnamefont {S.-P.}\ \bibnamefont
				{Chiu}}, \bibinfo {author} {\bibfnamefont {S.-S.}\ \bibnamefont {Yeh}},
			\bibinfo {author} {\bibfnamefont {C.-J.}\ \bibnamefont {Chiou}}, \bibinfo
			{author} {\bibfnamefont {Y.-C.}\ \bibnamefont {Chou}}, \bibinfo {author}
			{\bibfnamefont {J.-J.}\ \bibnamefont {Lin}}, \ and\ \bibinfo {author}
			{\bibfnamefont {C.-C.}\ \bibnamefont {Tsuei}},\ }\href {\doibase
			10.1021/acsnano.6b06553} {\bibfield  {journal} {\bibinfo  {journal} {ACS
					Nano}\ }\textbf {\bibinfo {volume} {11}},\ \bibinfo {pages} {516} (\bibinfo
			{year} {2017})}\BibitemShut {NoStop}%
		\bibitem [{\citenamefont {Gor'kov}\ and\ \citenamefont
			{Rashba}(2001)}]{GorkovRashba}%
		\BibitemOpen
		\bibfield  {author} {\bibinfo {author} {\bibfnamefont {L.~P.}\ \bibnamefont
				{Gor'kov}}\ and\ \bibinfo {author} {\bibfnamefont {E.~I.}\ \bibnamefont
				{Rashba}},\ }\href {\doibase 10.1103/PhysRevLett.87.037004} {\bibfield
			{journal} {\bibinfo  {journal} {Phys. Rev. Lett.}\ }\textbf {\bibinfo
				{volume} {87}},\ \bibinfo {pages} {037004} (\bibinfo {year}
			{2001})}\BibitemShut {NoStop}%
		\bibitem [{\citenamefont {Frigeri}\ \emph {et~al.}(2004)\citenamefont
			{Frigeri}, \citenamefont {Agterberg}, \citenamefont {Koga},\ and\
			\citenamefont {Sigrist}}]{Frigeri2004}%
		\BibitemOpen
		\bibfield  {author} {\bibinfo {author} {\bibfnamefont {P.~A.}\ \bibnamefont
				{Frigeri}}, \bibinfo {author} {\bibfnamefont {D.~F.}\ \bibnamefont
				{Agterberg}}, \bibinfo {author} {\bibfnamefont {A.}~\bibnamefont {Koga}}, \
			and\ \bibinfo {author} {\bibfnamefont {M.}~\bibnamefont {Sigrist}},\ }\href
		{\doibase 10.1103/PhysRevLett.92.097001} {\bibfield  {journal} {\bibinfo
				{journal} {Phys. Rev. Lett.}\ }\textbf {\bibinfo {volume} {92}},\ \bibinfo
			{pages} {097001} (\bibinfo {year} {2004})}\BibitemShut {NoStop}%
		\bibitem [{\citenamefont {Mattheiss}\ and\ \citenamefont
			{Hensel}(1989)}]{Mattheiss1989}%
		\BibitemOpen
		\bibfield  {author} {\bibinfo {author} {\bibfnamefont {L.~F.}\ \bibnamefont
				{Mattheiss}}\ and\ \bibinfo {author} {\bibfnamefont {J.~C.}\ \bibnamefont
				{Hensel}},\ }\href {\doibase 10.1103/PhysRevB.39.7754} {\bibfield  {journal}
			{\bibinfo  {journal} {Phys. Rev. B}\ }\textbf {\bibinfo {volume} {39}},\
			\bibinfo {pages} {7754} (\bibinfo {year} {1989})}\BibitemShut {NoStop}%
		\bibitem [{\citenamefont {Ekman}\ and\ \citenamefont {Ozoli\ifmmode
				\mbox{\c{n}}\else \c{n}\fi{}\ifmmode~\check{s}\else
				\v{s}\fi{}}(1998)}]{Ekman1998}%
		\BibitemOpen
		\bibfield  {author} {\bibinfo {author} {\bibfnamefont {M.}~\bibnamefont
				{Ekman}}\ and\ \bibinfo {author} {\bibfnamefont {V.}~\bibnamefont
				{Ozoli\ifmmode \mbox{\c{n}}\else \c{n}\fi{}\ifmmode~\check{s}\else
					\v{s}\fi{}}},\ }\href {\doibase 10.1103/PhysRevB.57.4419} {\bibfield
			{journal} {\bibinfo  {journal} {Phys. Rev. B}\ }\textbf {\bibinfo {volume}
				{57}},\ \bibinfo {pages} {4419} (\bibinfo {year} {1998})}\BibitemShut
		{NoStop}%
		\bibitem [{\citenamefont {Ma}\ and\ \citenamefont {Allen}(1994)}]{MaAllen1994}%
		\BibitemOpen
		\bibfield  {author} {\bibinfo {author} {\bibfnamefont {Z.}~\bibnamefont
				{Ma}}\ and\ \bibinfo {author} {\bibfnamefont {L.~H.}\ \bibnamefont {Allen}},\
		}\href {\doibase 10.1103/PhysRevB.49.13501} {\bibfield  {journal} {\bibinfo
				{journal} {Phys. Rev. B}\ }\textbf {\bibinfo {volume} {49}},\ \bibinfo
			{pages} {13501} (\bibinfo {year} {1994})}\BibitemShut {NoStop}%
		\bibitem [{Sup()}]{SupMat}%
		\BibitemOpen
		\href@noop {} {}\bibinfo {note} {see Supplemental Material.}\BibitemShut {Stop}%
		\bibitem [{\citenamefont {Yeh}\ \emph {et~al.}(2017)\citenamefont {Yeh},
			\citenamefont {Chang},\ and\ \citenamefont {Lin}}]{Yeh.17}%
		\BibitemOpen
		\bibfield  {author} {\bibinfo {author} {\bibfnamefont {S.-S.}\ \bibnamefont
				{Yeh}}, \bibinfo {author} {\bibfnamefont {W.-Y.}\ \bibnamefont {Chang}}, \
			and\ \bibinfo {author} {\bibfnamefont {J.-J.}\ \bibnamefont {Lin}},\ }\href
		{\doibase 10.1126/sciadv.1700135} {\bibfield  {journal} {\bibinfo  {journal}
				{Science Advances}\ }\textbf {\bibinfo {volume} {3}},\ \bibinfo {pages}
			{e1700135} (\bibinfo {year} {2017})}\BibitemShut {NoStop}%
		\bibitem [{\citenamefont {{Nazarov}}(1999)}]{Nazarov1999}%
		\BibitemOpen
		\bibfield  {author} {\bibinfo {author} {\bibfnamefont {Y.~V.}\ \bibnamefont
				{{Nazarov}}},\ }\href {\doibase 10.1006/spmi.1999.0738} {\bibfield  {journal}
			{\bibinfo  {journal} {Superlattices and Microstructures}\ }\textbf {\bibinfo
				{volume} {25}},\ \bibinfo {pages} {1221} (\bibinfo {year}
			{1999})}\BibitemShut {NoStop}%
		\bibitem [{\citenamefont {Tanaka}\ \emph {et~al.}(2003)\citenamefont {Tanaka},
			\citenamefont {Nazarov},\ and\ \citenamefont {Kashiwaya}}]{Tanaka2003bc}%
		\BibitemOpen
		\bibfield  {author} {\bibinfo {author} {\bibfnamefont {Y.}~\bibnamefont
				{Tanaka}}, \bibinfo {author} {\bibfnamefont {Y.~V.}\ \bibnamefont {Nazarov}},
			\ and\ \bibinfo {author} {\bibfnamefont {S.}~\bibnamefont {Kashiwaya}},\
		}\href {\doibase 10.1103/PhysRevLett.90.167003} {\bibfield  {journal}
			{\bibinfo  {journal} {Phys. Rev. Lett.}\ }\textbf {\bibinfo {volume} {90}},\
			\bibinfo {pages} {167003} (\bibinfo {year} {2003})}\BibitemShut {NoStop}%
		\bibitem [{\citenamefont {Tanaka}\ \emph
			{et~al.}(2004{\natexlab{a}})\citenamefont {Tanaka}, \citenamefont {Nazarov},
			\citenamefont {Golubov},\ and\ \citenamefont {Kashiwaya}}]{Tanaka2004}%
		\BibitemOpen
		\bibfield  {author} {\bibinfo {author} {\bibfnamefont {Y.}~\bibnamefont
				{Tanaka}}, \bibinfo {author} {\bibfnamefont {Y.~V.}\ \bibnamefont {Nazarov}},
			\bibinfo {author} {\bibfnamefont {A.~A.}\ \bibnamefont {Golubov}}, \ and\
			\bibinfo {author} {\bibfnamefont {S.}~\bibnamefont {Kashiwaya}},\ }\href
		{\doibase 10.1103/PhysRevB.69.144519} {\bibfield  {journal} {\bibinfo
				{journal} {Phys. Rev. B}\ }\textbf {\bibinfo {volume} {69}},\ \bibinfo
			{pages} {144519} (\bibinfo {year} {2004}{\natexlab{a}})}\BibitemShut
		{NoStop}%
		\bibitem [{\citenamefont {Tanaka}\ \emph
			{et~al.}(2004{\natexlab{b}})\citenamefont {Tanaka}, \citenamefont {Nazarov},
			\citenamefont {Golubov},\ and\ \citenamefont {Kashiwaya}}]{Tanaka2004_ERR}%
		\BibitemOpen
		\bibfield  {author} {\bibinfo {author} {\bibfnamefont {Y.}~\bibnamefont
				{Tanaka}}, \bibinfo {author} {\bibfnamefont {Y.~V.}\ \bibnamefont {Nazarov}},
			\bibinfo {author} {\bibfnamefont {A.~A.}\ \bibnamefont {Golubov}}, \ and\
			\bibinfo {author} {\bibfnamefont {S.}~\bibnamefont {Kashiwaya}},\ }\href
		{\doibase 10.1103/PhysRevB.70.219907} {\bibfield  {journal} {\bibinfo
				{journal} {Phys. Rev. B}\ }\textbf {\bibinfo {volume} {70}},\ \bibinfo
			{pages} {219907} (\bibinfo {year} {2004}{\natexlab{b}})}\BibitemShut
		{NoStop}%
		\bibitem [{\citenamefont {Lin}\ and\ \citenamefont {Bird}(2002)}]{Lin_2002}%
		\BibitemOpen
		\bibfield  {author} {\bibinfo {author} {\bibfnamefont {J.~J.}\ \bibnamefont
				{Lin}}\ and\ \bibinfo {author} {\bibfnamefont {J.~P.}\ \bibnamefont {Bird}},\
		}\href {\doibase 10.1088/0953-8984/14/18/201} {\bibfield  {journal} {\bibinfo
				{journal} {Journal of Physics: Condensed Matter}\ }\textbf {\bibinfo
				{volume} {14}},\ \bibinfo {pages} {R501} (\bibinfo {year}
			{2002})}\BibitemShut {NoStop}%
		\bibitem [{\citenamefont {Matthias}\ \emph {et~al.}(1980)\citenamefont
			{Matthias}, \citenamefont {Stewart}, \citenamefont {Giorgi}, \citenamefont
			{Smith}, \citenamefont {Fisk},\ and\ \citenamefont {Barz}}]{Mathhias1980}%
		\BibitemOpen
		\bibfield  {author} {\bibinfo {author} {\bibfnamefont {B.~T.}\ \bibnamefont
				{Matthias}}, \bibinfo {author} {\bibfnamefont {G.~R.}\ \bibnamefont
				{Stewart}}, \bibinfo {author} {\bibfnamefont {A.~L.}\ \bibnamefont {Giorgi}},
			\bibinfo {author} {\bibfnamefont {J.~L.}\ \bibnamefont {Smith}}, \bibinfo
			{author} {\bibfnamefont {Z.}~\bibnamefont {Fisk}}, \ and\ \bibinfo {author}
			{\bibfnamefont {H.}~\bibnamefont {Barz}},\ }\href {\doibase
			10.1126/science.208.4442.401} {\bibfield  {journal} {\bibinfo  {journal}
				{Science}\ }\textbf {\bibinfo {volume} {208}},\ \bibinfo {pages} {401}
			(\bibinfo {year} {1980})}\BibitemShut {NoStop}%
		\bibitem [{\citenamefont {Suhl}\ \emph {et~al.}(1980)\citenamefont {Suhl},
			\citenamefont {Matthias}, \citenamefont {Hecker},\ and\ \citenamefont
			{Smith}}]{Suhl1980}%
		\BibitemOpen
		\bibfield  {author} {\bibinfo {author} {\bibfnamefont {H.}~\bibnamefont
				{Suhl}}, \bibinfo {author} {\bibfnamefont {B.~T.}\ \bibnamefont {Matthias}},
			\bibinfo {author} {\bibfnamefont {S.}~\bibnamefont {Hecker}}, \ and\ \bibinfo
			{author} {\bibfnamefont {J.~L.}\ \bibnamefont {Smith}},\ }\href {\doibase
			10.1103/PhysRevLett.45.1707} {\bibfield  {journal} {\bibinfo  {journal}
				{Phys. Rev. Lett.}\ }\textbf {\bibinfo {volume} {45}},\ \bibinfo {pages}
			{1707} (\bibinfo {year} {1980})}\BibitemShut {NoStop}%
		\bibitem [{\citenamefont {Sigrist}(2009)}]{Sigrist_Review}%
		\BibitemOpen
		\bibfield  {author} {\bibinfo {author} {\bibfnamefont {M.}~\bibnamefont
				{Sigrist}},\ }\href {\doibase 10.1063/1.3225489} {\bibfield  {journal}
			{\bibinfo  {journal} {AIP Conference Proceedings}\ }\textbf {\bibinfo
				{volume} {1162}},\ \bibinfo {pages} {55} (\bibinfo {year}
			{2009})}\BibitemShut {NoStop}%
		\bibitem [{Note1()}]{Note1}%
		\BibitemOpen
		\bibinfo {note} {$\protect \mathaccentV {vec}17E{B}$ denotes the total
			magnetic field while $\protect \mathaccentV {vec}17E{H}$ refers to the
			external field as is common in the GL literature. $H$ has to be identified
			with $B$ of Figs.\ \ref {fig1} and \ref {fig2}.}\BibitemShut {Stop}%
		\bibitem [{\citenamefont {Maeno}\ \emph {et~al.}(1998)\citenamefont {Maeno},
			\citenamefont {Ando}, \citenamefont {Mori}, \citenamefont {Ohmichi},
			\citenamefont {Ikeda}, \citenamefont {NishiZaki},\ and\ \citenamefont
			{Nakatsuji}}]{Maeno1998}%
		\BibitemOpen
		\bibfield  {author} {\bibinfo {author} {\bibfnamefont {Y.}~\bibnamefont
				{Maeno}}, \bibinfo {author} {\bibfnamefont {T.}~\bibnamefont {Ando}},
			\bibinfo {author} {\bibfnamefont {Y.}~\bibnamefont {Mori}}, \bibinfo {author}
			{\bibfnamefont {E.}~\bibnamefont {Ohmichi}}, \bibinfo {author} {\bibfnamefont
				{S.}~\bibnamefont {Ikeda}}, \bibinfo {author} {\bibfnamefont
				{S.}~\bibnamefont {NishiZaki}}, \ and\ \bibinfo {author} {\bibfnamefont
				{S.}~\bibnamefont {Nakatsuji}},\ }\href {\doibase
			10.1103/PhysRevLett.81.3765} {\bibfield  {journal} {\bibinfo  {journal}
				{Phys. Rev. Lett.}\ }\textbf {\bibinfo {volume} {81}},\ \bibinfo {pages}
			{3765} (\bibinfo {year} {1998})}\BibitemShut {NoStop}%
		\bibitem [{\citenamefont {Ando}\ \emph {et~al.}(1999)\citenamefont {Ando},
			\citenamefont {Akima}, \citenamefont {Mori},\ and\ \citenamefont
			{Maeno}}]{Ando.99}%
		\BibitemOpen
		\bibfield  {author} {\bibinfo {author} {\bibfnamefont {T.}~\bibnamefont
				{Ando}}, \bibinfo {author} {\bibfnamefont {T.}~\bibnamefont {Akima}},
			\bibinfo {author} {\bibfnamefont {Y.}~\bibnamefont {Mori}}, \ and\ \bibinfo
			{author} {\bibfnamefont {Y.}~\bibnamefont {Maeno}},\ }\href {\doibase
			10.1143/JPSJ.68.1651} {\bibfield  {journal} {\bibinfo  {journal} {Journal of
					the Physical Society of Japan}\ }\textbf {\bibinfo {volume} {68}},\ \bibinfo
			{pages} {1651} (\bibinfo {year} {1999})}\BibitemShut {NoStop}%
		\bibitem [{\citenamefont {Mao}\ \emph {et~al.}(2001)\citenamefont {Mao},
			\citenamefont {Nelson}, \citenamefont {Jin}, \citenamefont {Liu},\ and\
			\citenamefont {Maeno}}]{Mao.01}%
		\BibitemOpen
		\bibfield  {author} {\bibinfo {author} {\bibfnamefont {Z.~Q.}\ \bibnamefont
				{Mao}}, \bibinfo {author} {\bibfnamefont {K.~D.}\ \bibnamefont {Nelson}},
			\bibinfo {author} {\bibfnamefont {R.}~\bibnamefont {Jin}}, \bibinfo {author}
			{\bibfnamefont {Y.}~\bibnamefont {Liu}}, \ and\ \bibinfo {author}
			{\bibfnamefont {Y.}~\bibnamefont {Maeno}},\ }\href {\doibase
			10.1103/PhysRevLett.87.037003} {\bibfield  {journal} {\bibinfo  {journal}
				{Phys. Rev. Lett.}\ }\textbf {\bibinfo {volume} {87}},\ \bibinfo {pages}
			{037003} (\bibinfo {year} {2001})}\BibitemShut {NoStop}%
		\bibitem [{\citenamefont {Wang}\ \emph {et~al.}(2017)\citenamefont {Wang},
			\citenamefont {Luo}, \citenamefont {Lou}, \citenamefont {Ortmann},
			\citenamefont {Mao}, \citenamefont {Liu},\ and\ \citenamefont
			{Wei}}]{Wang.17}%
		\BibitemOpen
		\bibfield  {author} {\bibinfo {author} {\bibfnamefont {H.}~\bibnamefont
				{Wang}}, \bibinfo {author} {\bibfnamefont {J.}~\bibnamefont {Luo}}, \bibinfo
			{author} {\bibfnamefont {W.}~\bibnamefont {Lou}}, \bibinfo {author}
			{\bibfnamefont {J.~E.}\ \bibnamefont {Ortmann}}, \bibinfo {author}
			{\bibfnamefont {Z.~Q.}\ \bibnamefont {Mao}}, \bibinfo {author} {\bibfnamefont
				{Y.}~\bibnamefont {Liu}}, \ and\ \bibinfo {author} {\bibfnamefont
				{J.}~\bibnamefont {Wei}},\ }\href {\doibase 10.1088/1367-2630/aa65c5}
		{\bibfield  {journal} {\bibinfo  {journal} {New Journal of Physics}\ }\textbf
			{\bibinfo {volume} {19}},\ \bibinfo {pages} {053001} (\bibinfo {year}
			{2017})}\BibitemShut {NoStop}%
		\bibitem [{\citenamefont {Sigrist}\ and\ \citenamefont
			{Monien}(2001)}]{Sigrist2001}%
		\BibitemOpen
		\bibfield  {author} {\bibinfo {author} {\bibfnamefont {M.}~\bibnamefont
				{Sigrist}}\ and\ \bibinfo {author} {\bibfnamefont {H.}~\bibnamefont
				{Monien}},\ }\href {\doibase 10.1143/JPSJ.70.2409} {\bibfield  {journal}
			{\bibinfo  {journal} {Journal of the Physical Society of Japan}\ }\textbf
			{\bibinfo {volume} {70}},\ \bibinfo {pages} {2409} (\bibinfo {year}
			{2001})}\BibitemShut {NoStop}%
		\bibitem [{\citenamefont {Kaneyasu}\ \emph {et~al.}(2019)\citenamefont
			{Kaneyasu}, \citenamefont {Enokida}, \citenamefont {Nomura}, \citenamefont
			{Hasegawa}, \citenamefont {Sakai},\ and\ \citenamefont
			{Sigrist}}]{Sigrist2019}%
		\BibitemOpen
		\bibfield  {author} {\bibinfo {author} {\bibfnamefont {H.}~\bibnamefont
				{Kaneyasu}}, \bibinfo {author} {\bibfnamefont {Y.}~\bibnamefont {Enokida}},
			\bibinfo {author} {\bibfnamefont {T.}~\bibnamefont {Nomura}}, \bibinfo
			{author} {\bibfnamefont {Y.}~\bibnamefont {Hasegawa}}, \bibinfo {author}
			{\bibfnamefont {T.}~\bibnamefont {Sakai}}, \ and\ \bibinfo {author}
			{\bibfnamefont {M.}~\bibnamefont {Sigrist}},\ }\href {\doibase
			10.1103/PhysRevB.100.214501} {\bibfield  {journal} {\bibinfo  {journal}
				{Phys. Rev. B}\ }\textbf {\bibinfo {volume} {100}},\ \bibinfo {pages}
			{214501} (\bibinfo {year} {2019})}\BibitemShut {NoStop}%
		\bibitem [{\citenamefont {De~Gennes}(1964)}]{DG1964}%
		\BibitemOpen
		\bibfield  {author} {\bibinfo {author} {\bibfnamefont {P.~G.}\ \bibnamefont
				{De~Gennes}},\ }\href {\doibase 10.1103/RevModPhys.36.225} {\bibfield
			{journal} {\bibinfo  {journal} {Rev. Mod. Phys.}\ }\textbf {\bibinfo {volume}
				{36}},\ \bibinfo {pages} {225} (\bibinfo {year} {1964})}\BibitemShut
		{NoStop}%
		\bibitem [{\citenamefont {Tanaka}\ \emph {et~al.}(2012)\citenamefont {Tanaka},
			\citenamefont {Sato},\ and\ \citenamefont {Nagaosa}}]{TSN_Review}%
		\BibitemOpen
		\bibfield  {author} {\bibinfo {author} {\bibfnamefont {Y.}~\bibnamefont
				{Tanaka}}, \bibinfo {author} {\bibfnamefont {M.}~\bibnamefont {Sato}}, \ and\
			\bibinfo {author} {\bibfnamefont {N.}~\bibnamefont {Nagaosa}},\ }\href
		{\doibase 10.1143/JPSJ.81.011013} {\bibfield  {journal} {\bibinfo  {journal}
				{Journal of the Physical Society of Japan}\ }\textbf {\bibinfo {volume}
				{81}},\ \bibinfo {pages} {011013} (\bibinfo {year} {2012})}\BibitemShut
		{NoStop}%
		\bibitem [{\citenamefont {Linder}\ and\ \citenamefont
			{Balatsky}(2019)}]{Lindner.19}%
		\BibitemOpen
		\bibfield  {author} {\bibinfo {author} {\bibfnamefont {J.}~\bibnamefont
				{Linder}}\ and\ \bibinfo {author} {\bibfnamefont {A.~V.}\ \bibnamefont
				{Balatsky}},\ }\href {\doibase 10.1103/RevModPhys.91.045005} {\bibfield
			{journal} {\bibinfo  {journal} {Rev. Mod. Phys.}\ }\textbf {\bibinfo {volume}
				{91}},\ \bibinfo {pages} {045005} (\bibinfo {year} {2019})}\BibitemShut
		{NoStop}%
		\bibitem [{\citenamefont {Ando}\ \emph {et~al.}(2020)\citenamefont {Ando},
			\citenamefont {Miyasaka}, \citenamefont {Li}, \citenamefont {Ishizuka},
			\citenamefont {Arakawa}, \citenamefont {Shiota}, \citenamefont {Moriyama},
			\citenamefont {Yanase},\ and\ \citenamefont {Ono}}]{Ando.20}%
		\BibitemOpen
		\bibfield  {author} {\bibinfo {author} {\bibfnamefont {F.}~\bibnamefont
				{Ando}}, \bibinfo {author} {\bibfnamefont {Y.}~\bibnamefont {Miyasaka}},
			\bibinfo {author} {\bibfnamefont {T.}~\bibnamefont {Li}}, \bibinfo {author}
			{\bibfnamefont {J.}~\bibnamefont {Ishizuka}}, \bibinfo {author}
			{\bibfnamefont {T.}~\bibnamefont {Arakawa}}, \bibinfo {author} {\bibfnamefont
				{Y.}~\bibnamefont {Shiota}}, \bibinfo {author} {\bibfnamefont
				{T.}~\bibnamefont {Moriyama}}, \bibinfo {author} {\bibfnamefont
				{Y.}~\bibnamefont {Yanase}}, \ and\ \bibinfo {author} {\bibfnamefont
				{T.}~\bibnamefont {Ono}},\ }\href {\doibase 10.1038/s41586-020-2590-4}
		{\bibfield  {journal} {\bibinfo  {journal} {Nature}\ }\textbf {\bibinfo
				{volume} {584}},\ \bibinfo {pages} {373–376} (\bibinfo {year}
			{2020})}\BibitemShut {NoStop}%
		\bibitem [{\citenamefont {Wakatsuki}\ \emph {et~al.}(2017)\citenamefont
			{Wakatsuki}, \citenamefont {Saito}, \citenamefont {Hoshino}, \citenamefont
			{Itahashi}, \citenamefont {Ideue}, \citenamefont {Ezawa}, \citenamefont
			{Iwasa},\ and\ \citenamefont {Nagaosa}}]{Wakatsuki.17}%
		\BibitemOpen
		\bibfield  {author} {\bibinfo {author} {\bibfnamefont {R.}~\bibnamefont
				{Wakatsuki}}, \bibinfo {author} {\bibfnamefont {Y.}~\bibnamefont {Saito}},
			\bibinfo {author} {\bibfnamefont {S.}~\bibnamefont {Hoshino}}, \bibinfo
			{author} {\bibfnamefont {Y.~M.}\ \bibnamefont {Itahashi}}, \bibinfo {author}
			{\bibfnamefont {T.}~\bibnamefont {Ideue}}, \bibinfo {author} {\bibfnamefont
				{M.}~\bibnamefont {Ezawa}}, \bibinfo {author} {\bibfnamefont
				{Y.}~\bibnamefont {Iwasa}}, \ and\ \bibinfo {author} {\bibfnamefont
				{N.}~\bibnamefont {Nagaosa}},\ }\href {\doibase 10.1126/sciadv.1602390}
		{\bibfield  {journal} {\bibinfo  {journal} {Science Advances}\ }\textbf
			{\bibinfo {volume} {3}},\ \bibinfo {pages} {e1602390} (\bibinfo {year}
			{2017})}\BibitemShut {NoStop}%
	\end{thebibliography}

\begin{thebibliography}{5}%
		\makeatletter
		\providecommand \@ifxundefined [1]{%
			\@ifx{#1\undefined}
		}%
		\providecommand \@ifnum [1]{%
			\ifnum #1\expandafter \@firstoftwo
			\else \expandafter \@secondoftwo
			\fi
		}%
		\providecommand \@ifx [1]{%
			\ifx #1\expandafter \@firstoftwo
			\else \expandafter \@secondoftwo
			\fi
		}%
		\providecommand \natexlab [1]{#1}%
		\providecommand \enquote  [1]{``#1''}%
		\providecommand \bibnamefont  [1]{#1}%
		\providecommand \bibfnamefont [1]{#1}%
		\providecommand \citenamefont [1]{#1}%
		\providecommand \href@noop [0]{\@secondoftwo}%
		\providecommand \href [0]{\begingroup \@sanitize@url \@href}%
		\providecommand \@href[1]{\@@startlink{#1}\@@href}%
		\providecommand \@@href[1]{\endgroup#1\@@endlink}%
		\providecommand \@sanitize@url [0]{\catcode `\\12\catcode `\$12\catcode
			`\&12\catcode `\#12\catcode `\^12\catcode `\_12\catcode `\%12\relax}%
		\providecommand \@@startlink[1]{}%
		\providecommand \@@endlink[0]{}%
		\providecommand \url  [0]{\begingroup\@sanitize@url \@url }%
		\providecommand \@url [1]{\endgroup\@href {#1}{\urlprefix }}%
		\providecommand \urlprefix  [0]{URL }%
		\providecommand \Eprint [0]{\href }%
		\providecommand \doibase [0]{http://dx.doi.org/}%
		\providecommand \selectlanguage [0]{\@gobble}%
		\providecommand \bibinfo  [0]{\@secondoftwo}%
		\providecommand \bibfield  [0]{\@secondoftwo}%
		\providecommand \translation [1]{[#1]}%
		\providecommand \BibitemOpen [0]{}%
		\providecommand \bibitemStop [0]{}%
		\providecommand \bibitemNoStop [0]{.\EOS\space}%
		\providecommand \EOS [0]{\spacefactor3000\relax}%
		\providecommand \BibitemShut  [1]{\csname bibitem#1\endcsname}%
		\let\auto@bib@innerbib\@empty
		\bibitem [{\citenamefont {Sigrist}(2009)}]{SSigrist_Review}%
		\BibitemOpen
		\bibfield  {author} {\bibinfo {author} {\bibfnamefont {M.}~\bibnamefont
				{Sigrist}},\ }\href {\doibase 10.1063/1.3225489} {\bibfield  {journal}
			{\bibinfo  {journal} {AIP Conference Proceedings}\ }\textbf {\bibinfo
				{volume} {1162}},\ \bibinfo {pages} {55} (\bibinfo {year}
			{2009})}\BibitemShut {NoStop}%
		\bibitem [{\citenamefont {Ma}\ \emph {et~al.}(1991)\citenamefont {Ma},
			\citenamefont {Allen},\ and\ \citenamefont {Lee}}]{SMa.91}%
		\BibitemOpen
		\bibfield  {author} {\bibinfo {author} {\bibfnamefont {Z.}~\bibnamefont
				{Ma}}, \bibinfo {author} {\bibfnamefont {L.~H.}\ \bibnamefont {Allen}}, \
			and\ \bibinfo {author} {\bibfnamefont {S.}~\bibnamefont {Lee}},\ }\href
		{\doibase 10.1557/PROC-237-661} {\bibfield  {journal} {\bibinfo  {journal}
				{MRS Proceedings}\ }\textbf {\bibinfo {volume} {237}},\ \bibinfo {pages}
			{661} (\bibinfo {year} {1991})}\BibitemShut {NoStop}%
		\bibitem [{\citenamefont {Chen}(2004)}]{SChen.04}%
		\BibitemOpen
		\bibfield  {author} {\bibinfo {author} {\bibfnamefont {L.~J.}\ \bibnamefont
				{Chen}},\ }\href@noop {} {\emph {\bibinfo {title} {Silicide Technology for
					Integrated Circuits}}},\ EMIS Processing Series\ (\bibinfo  {publisher} {The
			Institution of Engineering and Technology},\ \bibinfo {year}
		{2004})\BibitemShut {NoStop}%
		\bibitem [{\citenamefont {Chiu}\ \emph {et~al.}(2021)\citenamefont {Chiu},
			\citenamefont {Tsuei}, \citenamefont {Yeh}, \citenamefont {Zhang},
			\citenamefont {Kirchner},\ and\ \citenamefont {Lin}}]{SChiu.21}%
		\BibitemOpen
		\bibfield  {author} {\bibinfo {author} {\bibfnamefont {S.-P.}\ \bibnamefont
				{Chiu}}, \bibinfo {author} {\bibfnamefont {C.~C.}\ \bibnamefont {Tsuei}},
			\bibinfo {author} {\bibfnamefont {S.-S.}\ \bibnamefont {Yeh}}, \bibinfo
			{author} {\bibfnamefont {F.-C.}\ \bibnamefont {Zhang}}, \bibinfo {author}
			{\bibfnamefont {S.}~\bibnamefont {Kirchner}}, \ and\ \bibinfo {author}
			{\bibfnamefont {J.-J.}\ \bibnamefont {Lin}},\ }\href {\doibase
			10.1126/sciadv.abg6569} {\bibfield  {journal} {\bibinfo  {journal} {Science
					Advances}\ }\textbf {\bibinfo {volume} {7}},\ \bibinfo {pages} {eabg6569}
			(\bibinfo {year} {2021})}\BibitemShut {NoStop}%
		\bibitem [{\citenamefont {Mammoliti}\ \emph {et~al.}(2002)\citenamefont
			{Mammoliti}, \citenamefont {Grimaldi},\ and\ \citenamefont
			{La~Via}}]{SMammoliti.02}%
		\BibitemOpen
		\bibfield  {author} {\bibinfo {author} {\bibfnamefont {F.}~\bibnamefont
				{Mammoliti}}, \bibinfo {author} {\bibfnamefont {M.~G.}\ \bibnamefont
				{Grimaldi}}, \ and\ \bibinfo {author} {\bibfnamefont {F.}~\bibnamefont
				{La~Via}},\ }\href {\doibase 10.1063/1.1500787} {\bibfield  {journal}
			{\bibinfo  {journal} {J. Appl. Phys.}\ }\textbf {\bibinfo {volume} {92}},\
			\bibinfo {pages} {3147} (\bibinfo {year} {2002})}\BibitemShut {NoStop}%
	\end{thebibliography}
\end{document}